\documentclass[aip,pof,11pt]{revtex4-1}
\usepackage{graphicx}
\usepackage{color}
\usepackage{import}
\usepackage{amsmath}
\usepackage{multirow}
\usepackage{bigdelim}
\usepackage{epstopdf}
\usepackage{textcomp}
\usepackage{placeins}
\usepackage{pbox}

\newcommand{\ts}[2]{\ensuremath{\mathrm{#1}_{\mathrm{#2}}}} 
\newcommand{\Figref}[1]{Figure~\ref{#1}}

\newcommand{\Eqnref}[1]{Equation~\ref{#1}}
\newcommand{\Secref}[1]{Section~\ref{#1}}

\newcommand{\figref}[1]{Fig.~\ref{#1}}
\newcommand{\tabref}[1]{Table~\ref{#1}}
\newcommand{\eqnref}[1]{Eq.~\ref{#1}}
\newcommand{\secref}[1]{Sec.~\ref{#1}}
\newcommand{\appref}[1]{App.~\ref{#1}}
\begin{document}

% Use the \preprint command to place your local institutional report number 
% on the title page in preprint mode.
% Multiple \preprint commands are allowed.
%\preprint{}

\title{The impact of heating the breakdown bubble on the global mode of a swirling jet: Experiments and linear stability analysis.} %Title of paper

% repeat the \author .. \affiliation  etc. as needed
% \email, \thanks, \homepage, \altaffiliation all apply to the current author.
% Explanatory text should go in the []'s, 
% actual e-mail address or url should go in the {}'s for \email and \homepage.
% Please use the appropriate macro for the type of information

% \affiliation command applies to all authors since the last \affiliation command. 
% The \affiliation command should follow the other information.

\author{Lothar Rukes}
\email[]{lothar.rukes@tu-berlin.de}
%\homepage[]{Your web page}
%\thanks{}
%\altaffiliation{}
\affiliation{Chair of Fluid Dynamics, Hermann-F\"ottinger-Institut, Technische Universit\"at Berlin}

\author{Moritz Sieber}
%\homepage[]{Your web page}
%\thanks{}
%\altaffiliation{}
\affiliation{Chair of Fluid Dynamics, Hermann-F\"ottinger-Institut, Technische Universit\"at Berlin}

\author{C. Oliver Paschereit}
%\homepage[]{Your web page}
%\thanks{}
%\altaffiliation{}
\affiliation{Chair of Fluid Dynamics, Hermann-F\"ottinger-Institut, Technische Universit\"at Berlin}

\author{Kilian Oberleithner}
%\homepage[]{Your web page}
%\thanks{}
%\altaffiliation{}
\affiliation{Chair of Fluid Dynamics, Hermann-F\"ottinger-Institut, Technische Universit\"at Berlin}

% Collaboration name, if desired (requires use of superscriptaddress option in \documentclass). 
% \noaffiliation is required (may also be used with the \author command).
%\collaboration{}
%\noaffiliation

\date{\today}

\begin{abstract}
This study investigates the dynamics of non-isothermal swirling jets undergoing vortex breakdown, with an emphasis on helical coherent structures. It is proposed that the dominant helical coherent structure can be suppressed by heating the recirculation bubble. This proposition is assessed with Stereo Particle Image Velocimetry (PIV) measurements of the breakdown region of isothermal and heated swirling jets. The coherent kinetic energy of the dominant helical structure was derived from PIV snapshots via Proper Orthogonal Decomposition. For one set of experimental parameters, mild heating is found to increase the energy content of the dominant helical mode. Strong heating leads to a reduction by 30\% of the coherent structures energy. For a second set of experimental parameters, no alteration of the dominant coherent structure is detectable. Local linear stability analysis of the time-averaged velocity fields shows that the key difference between the two configurations is the density ratio at the respective wavemaker location. A density ratio of approximately 0.8 is found to correlate to a suppression of the global mode in the experiments. A parametric study with model density and velocity profiles indicates the most important parameters that govern the local absolute growth rate: The density ratio and the relative position of the density profiles and the inner shear layer.    
    
\end{abstract}

% insert suggested PACS numbers in braces on next line

\maketitle %\maketitle must follow title, authors, abstract and \pacs

% Body of paper goes here. Use proper sectioning commands. 
% References should be done using the \cite, \ref, and \label commands
%-------------------------------------------------------------------------------------------------------------------------------------------------------------------------------------------------------------------
%-------------------------------------------------------------------------------------------------------------------------------------------------------------------------------------------------------------------
\section{Introduction}
\label{sec:Introduction}
\begin{figure}
\includegraphics[width = 0.75\textwidth]{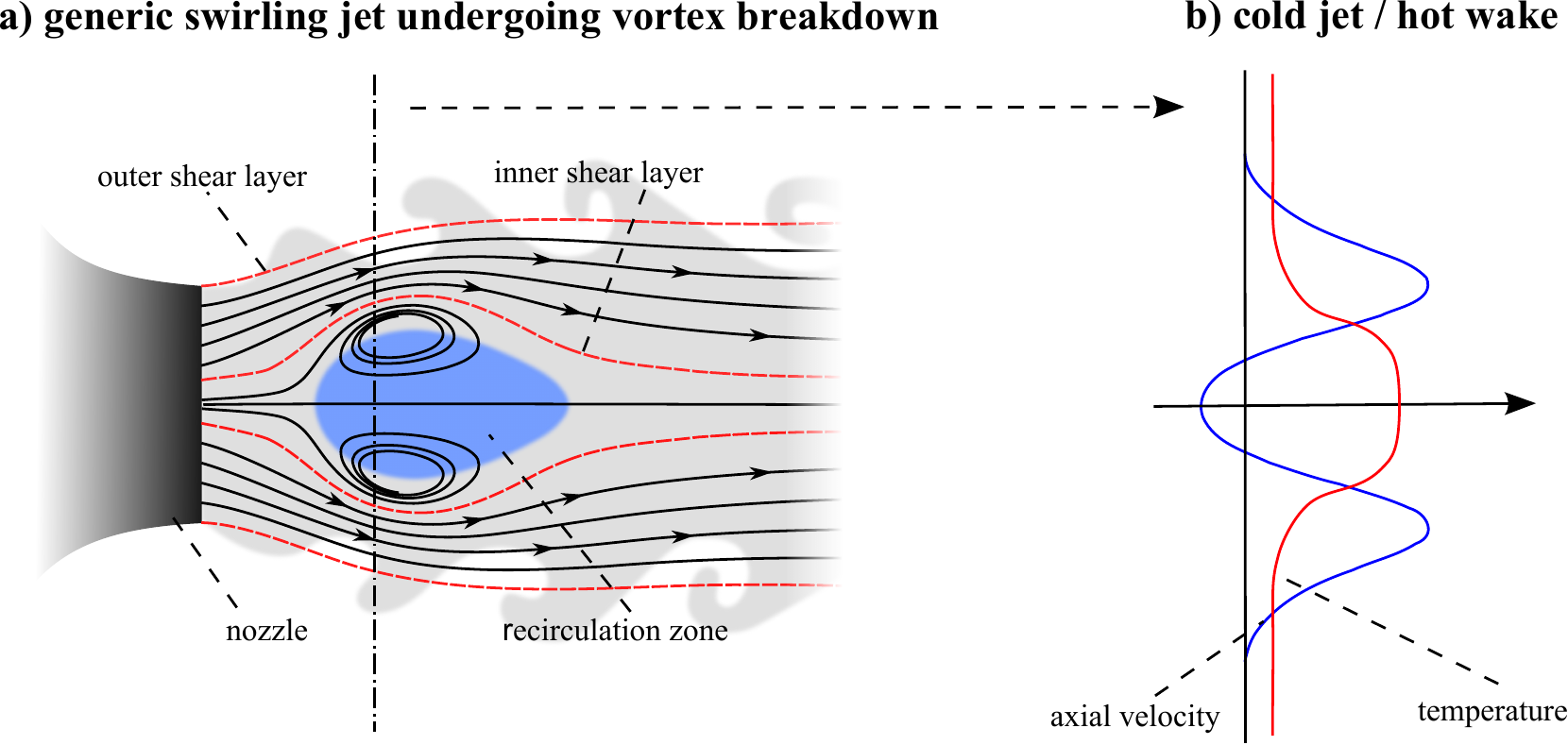}%
\caption{Sketch of a swirling jet undergoing vortex breakdown with superimposed temperature distribution.\label{fig:Sketch}}%
\end{figure}

The vast majority of modern gas turbine burners use swirling jets undergoing vortex breakdown as a means of stabilizing the flame position \cite{LuccaNegro.2001,Huang.2009}. Measurements of the isothermal flow consistently show the presence of a large-scale coherent structure, the global mode, as soon as vortex breakdown is present in the mean flow \cite{Liang.2005,Oberleithner.2010,Billant.1998}. However, measurements of the reacting flow revealed that the global mode may be suppressed, depending on the position of the flame relative to the flow \cite{Terhaar.2014b}. This is remarkable, since the global mode is typically a robust feature of the flow that is difficult to control in general.\\
In recent years, significant progress was made in analyzing the global mode in isothermal swirling jets. \textcolor{black}{Key contributions to our understanding of the flow physics were made by the transient swirl experiments of Liang \& Maxworthy \cite{Liang.2005}, the description of vortex breakdown types \cite{Billant.1998}, the interpretation of the spiral vortex breakdown mode as a global stability mode \cite{Ruith.2003b,Gallaire.2006,Gallaire.2003b} and the analysis of fundamental stability properties of swirling flows\cite{Gallaire.2003,Loiseleux.1998,Parras.2007,Olendraru.1999}}. The swirl number was identified as an important control parameter describing the evolution of swirling jets. According to the study of Oberleithner et al. \cite{Oberleithner.2012}, two distinct swirl numbers can be identified that separate three regimes of different dynamics. At values of the swirl number below $\ts{S}{VB}$, either no recirculation zone or an intermittently occurring recirculation zone is present in the flow. At swirl levels above $\ts{S}{VB}$, a zone of recirculating fluid is permanently present in the flow. A global mode is only present in the flow, if the swirl number is above a critical level $\ts{S}{crit}$, with $\ts{S}{crit} > \ts{S}{VB}$. At swirl numbers larger than $\ts{S}{crit}$, the flow was observed to oscillate on a limit cycle.\\
The global mode was recognized as a helical coherent structure that winds around the recirculation zone and is observed to have an azimuthal wave number of one, co-rotating with the base flow \cite{Oberleithner.2011b,Ruith.2003b} \textcolor{black}{in time but winding opposite to the swirl in space}. Concerning the temporal dynamics, the global mode is known to oscillate at a defined frequency. The global mode is consistently observed to dominate the entire dynamics of swirling flows undergoing vortex breakdown once it is present \cite{Oberleithner.2011b,Ruith.2003b,Terhaar.2014,Liang.2005}. \\
While the dynamical behavior of the global mode is well established for isothermal jets, the situation is less clear in non-isothermal jets. Experimental and numerical investigations of an industrial-type swirl-stabilized burner showed that the global mode was present in the isothermal flow, whereas it was not present in the reacting flow \cite{Roux.2005,Selle.2004}. In contrast, in the studies of Syred et al. \cite{Syred.2006,Syred.1997} the global mode was present for all burner configurations considered. Terhaar et al. \cite{Terhaar.2014b} conducted PIV measurements of the reacting flow in a premixed swirl-stabilized burner and observed that the global mode is suppressed, if the flame front is located close to the burner inlet. No effect on the global mode was detected, when the flame was located further downstream in the burner. While this indicates that the temperature gradient produced by the flame plays an important role in the suppression of the global mode, the flow in a gas turbine combustor is influenced by many aspects. Syred \cite{Syred.2006} summarizes that the occurrence of the global mode in reacting burner flows is not only dependent on the presence of vortex breakdown, but also on the burner geometry, the confinement, the equivalence ratio and on the mode of fuel injection. \\
Whether isothermal or non-isothermal swirling jets are considered, local linear stability analysis has proven to be a useful tool in the analysis of such flows. The cornerstone of the analysis is the distinction between convectively and absolutely unstable flows. Gallaire et al. \cite{Gallaire.2006} showed that swirling jets are absolutely unstable when vortex breakdown is present. They demonstrated that the global mode observed in experiments can be interpreted as a global stability mode. The connection from the local concept of absolute instability to the global mode is made via a frequency selection criterion. This criterion yields the location of the so-called wavemaker, at which the frequency and growth rate of the global mode are determined. \\
While stability analysis is strictly applicable only to a base flow that is a solution of the stationary Navier-Stokes equations, it was recently applied to time-averaged flows with great success. Barkley \cite{Barkley.2006} and Leontini et al.\cite{LEONTINI.2010} studied the capability of (global) linear stability analysis in predicting the frequency of the vortex shedding in the mean wake flow of a circular cylinder. They found an excellent agreement with the frequency derived from nonlinear direct numerical simulation. Pier \cite{Pier.2002} evaluated the ability of several frequency selection criteria to predict the frequency of vortex shedding in the cylinder wake within the WKBJ framework. The criterion for a linear global mode, as introduced by Chomaz et al.\cite{Chomaz.1991}, showed the best agreement with the nonlinear direct numerical simulation. Oberleithner et al. \cite{Oberleithner.2011b} calculated the three-dimensional shape of the global mode from eigenfunctions of a local mean flow stability analysis. They found a good agreement between their stability calculations and the shape of the global mode derived from experimental data. Oberleithner et al. \cite{Oberleithner.2014b} provide a discussion of stability analysis applied to the mean flow of a forced laminar jet. They found that the stability analysis accurately predicts the spatial growth and decay rate, phase velocity, as well as the phase and amplitude distribution of the excited coherent structure.  \\
Local stability analysis is only valid for weakly non-parallel flows. In swirling jets undergoing vortex breakdown, this is a strong assumption, since the recirculation bubble forces the jet to a wide expansion. However, Gallaire et al. \cite{Gallaire.2006} found a good agreement between the frequency of the global mode computed from local stability analysis and that derived from a direct numerical simulation. Similar good agreements between local stability analysis and simulation, respectively experiment were achieved by Oberleithner et al. \cite{Oberleithner.2011b}, Terhaar et al. \cite{Terhaar.2015}, Juniper et al. \cite{Juniper.2011} and Thiria \& Wesfreid \cite{Thiria.2007}\\
While the above cited studies all applied stability analysis to isothermal swirling jets, Oberleithner et al. \cite{Oberleithner.2013, Oberleithner.2015c} analyzed the non-isothermal flow field of a swirl flame. They considered two different flame shapes, a detached M-flame and a attached V-flame, where only the first featured a global mode. They performed a  local linear stability analysis of both flames based on measured mean velocity and density fields. For the M-flame, they identified the wavemaker near the upstream end of the recirculation bubble and could accurately predict the global mode frequency. For the V-flame, they noticed a strong density stratification in the wavemaker region, which leads to the suppression of the global mode. \\
Manoharan et al. \cite{Manoharan.2015} conducted a stability analysis of velocity and density profiles typically found in swirl combustors. These authors found that the presence of a density gradient reduces the extent of the region of absolute instability in the parameter space they considered. A similar conclusion is reached in the model study of Terhaar et al. \cite{Terhaar.2015}. Emerson et al. \cite{Emerson.2012} investigated the non-isothermal flows in bluff-body stabilized combustors without swirl. Their stability analysis revealed that the relative position of the density and velocity gradients is crucial for the large-scale dynamics of the wake flow.
\\ 
In combustion experiments, the density field is always connected to the flame shape and cannot be adjusted freely. Furthermore, in combustion experiments only two extreme cases can be considered, where either no density gradient is present at isothermal conditions or a very strong density gradient prevails at combustion conditions. Temperature measurements in combustion experiments are problematic and suffer from low spatial resolution \cite{Goke.2013}, low accuracy \cite{Hindasageri.2013} or require substantial density gradients to work reasonable \cite{Terhaar.2015b}. The investigation of intermediate heating regimes is not possible in these setups. We thus propose to study the impact of a density field in a more controlled environment that emulates the thermal configuration in the combustion experiments and facilitates temperature and velocity measurements. As shown in \figref{fig:Sketch} b), we investigate the global mode in a heated swirling jet undergoing vortex breakdown, where the heating is applied only in the breakdown bubble. This allows for a continuous adjustment of the relevant density gradients without changing the flow morphology. With this experimental approach, we attempt to contribute to the following open questions:
\begin{itemize}
\item
\textit{Is the suppression of the global mode in non-isothermal swirling jets an "on-off'' transition, or do intermediate states exist?}
\item
\textit{Is the amplitude reduction of the global mode achieved by the entire density field, or are there regions of particular sensitivity?}
\item
\textit{Swirling jets feature two axial shear layers that promote instability. Does the density-velocity collocation of both equally contribute to the global mode?}
\end{itemize}

The paper is organized as follows.  \Secref{sec:1} introduces the experimental facility, the important control parameters, the data acquisition and evaluation procedure and a discussion of the measurement error. The POD methodology and the stability analysis approach are introduced in \secref{sec:2}. The time-averaged temperature and velocity fields are presented in \secref{sec:3}, together with a discussion of the sole influence of the heating element on the flow. This section further includes an analysis of the energy content of the global mode, the results of the stability analysis applied to the measured data and a parametric study.

%-------------------------------------------------------------------------------------------------------------------------------------------------------------------------------------------------------------------
%------------------------------------------------------------------------------------------------------------------------------------------------------------------------------------------------------------------- 
\section{Experimental Setup}
\label{sec:1}
%----------------------------------------------------------------------------------------------------------------------------------------------------------------------------------------------------------------
\subsection{The swirling jet test rig}
\begin{figure}
\includegraphics{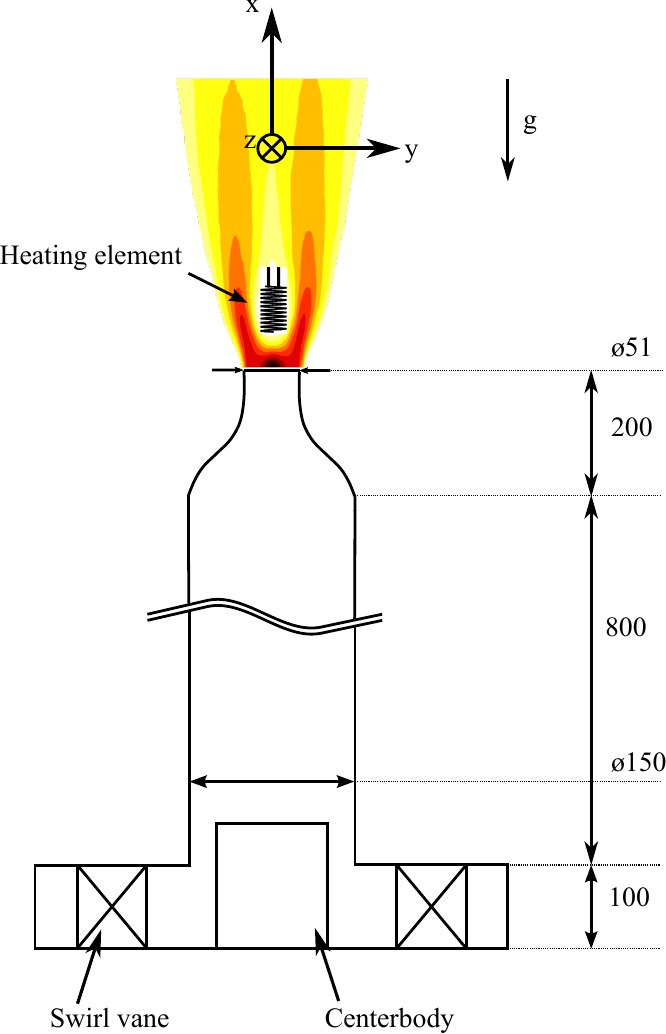}%
\caption{The swirling jet facility. All dimensions are in millimeters (not to scale).\label{fig:ExpSetup}}%
\end{figure}

The experimental setup used for the present investigation is shown in \figref{fig:ExpSetup}. It consist of a swirler with eleven continuously adjustable vanes, a pipe followed by a nozzle and the unconfined free field. The nozzle has an exit diameter of 51 mm. A center-body is mounted in the upstream confined part of the flow, in order to stabilize the position of the breakdown bubble in the unconfined flow \cite{Rukes.2015}. Flow measurements are obtained in the unconfined part of the setup. The swirler is fed from a pressurized air source via a mass flow controller. \\
\Figref{fig:ExpSetup} also indicates the presence of an electrical heating element in the flow. It is connected to a power supply unit that provides the heating current. At the largest heating current considered, the heating element is supplied with 174 Watt of electrical power. The heating element is placed inside the recirculation bubble. This ensures that heated fluid is recirculated back to the upstream stagnation point. It should be emphasized that the heating element in the sketch of \figref{fig:ExpSetup} is not drawn to scale and that the heating element is not a solid body. It consists of a curled heating wire with a thickness of 0.6 mm. The heating element has the form of a cylinder with a radius of 7 mm and a height of 50 mm. The influence of the heating element on the flow will be discussed separately in \secref{sec:HE}.\\
\autoref{fig:ExpSetup} also introduces a cartesian coordinate system with its origin placed on the jet axis in the nozzle exit plane. The $x$-axis is oriented in the flow direction, the $y$-axis in the cross-flow direction and the $z$-axis in the out-of-plane direction. Additionally, a cylindrical coordinate system is introduced with the same origin. The $r$-axis is aligned with the $y$-axis of the cartesian system at zero degrees of revolution, $\theta$ is counted mathematically positive and $x$ points in the streamwise direction. The velocity vector $\textbf{v}$ has the components $v_{x}$, $v_{y}$ and $v_{z}$ in the cartesian system, and $v_{r}$, $v_{\theta}$, $v_{z}$ in the polar coordinate system.
%----------------------------------------------------------------------------------------------------------------------------------------------------------------------------------------------------------------
\subsection{Operating conditions}
The heated swirling jet is characterized by the Reynolds number, 
\begin{align}
\mathrm{Re} = \frac{v_{bulk} \mathrm{D}}{\nu}, %\\
%\intertext{with the bulk velocity}
%v_{bulk} &= \frac{Q}{\pi (\mathrm{D}/2)^{2}}.\nonumber
\end{align}
with the bulk velocity $v_{bulk} = \frac{Q}{\pi (\mathrm{D}/2)^{2}}$. $Q$ denotes the volumetric flow, D the nozzle diameter and $\nu$ the kinematic viscosity of air. For all considered operating conditions the Mach number, $\mathrm{Ma} = \frac{v_{bulk}}{\mathrm{c}}$, is of the order of 0.01 and the flow is considered as incompressible, with c denoting the speed of sound. The swirl intensity is quantified at the nozzle lip with the swirl number \cite{Chigier.1967}
\begin{equation}
\mathrm{S} = \frac{\dot{G_{\theta}}}{\mathrm{D}/2  \dot{G_{x}}} = \frac{2\pi  \int\limits_{0}^{\infty}\rho  \overline{v}_{x} \overline{v}_{\theta}  r^{2} \mathrm{d}r}{\textrm{D}  \pi \int\limits_{0}^{\infty}\rho \left(\overline{v_{x}^{2}}- \frac{\overline{v_{\theta}^{2}}}{2}\right)  r \mathrm{d}r}.
\end{equation}
\textcolor{black}{$\dot{G_{\theta}}$ denotes the axial flux of azimuthal momentum and $\dot{G_{x}}$ denotes the axial flux of axial momentum.} All swirl numbers considered in this work are above critical \cite{Oberleithner.2012} and vortex breakdown is present in the mean field and the global mode can be expected to oscillate on its limit cycle. The heating is quantified by the density ratio between the jet axis at $r = 0$ and the ambient density,
\begin{equation}
\rho^{*} = \frac{\overline{\rho}_{cl}}{\rho_{\infty}}.
\end{equation}
The stronger the heating, the closer $\rho^{*}$ is to zero. The Richardson number characterizes the influence of gravity on the flow and is given by \cite{Monkewitz.1990}
\begin{equation}
\mathrm{Ri} = \mathrm{g}\frac{1-\rho^{*}}{\rho^{*}\overline{v}_{x \mathrm{max}}^{2}}\ts{x}{st}, 
\end{equation}
with g as the gravitational acceleration. The original definition of the Richardson number introduced in Monkewitz et al. \cite{Monkewitz.1990} is based on characteristic quantities in the nozzle exit plane. These authors considered hot jets, where the largest density ratio is attained in the nozzle exit plane. In the present study heat is added to the flow downstream of the nozzle, rendering the jet in the nozzle exit plane isothermal, and hence, the original definition may be misleading. Instead, the Richardson number is defined in terms of the axial location of the upstream stagnation point, $\ts{x}{st}$, and the velocity $v_{x \mathrm{max}} = \max(v_{x}(r,\ts{x}{VB}))$. Since the Richardson number is in any case markedly smaller than one, the influence of gravity is neglected \cite{Papanicolaou.1988,Monkewitz.1990}.\\
The values of all characteristic numbers are tabulated in \tabref{tab:ExpPara}. This table introduces the flow configurations that are investigated in the following. To reference each configuration, a shorthand notation is introduced, reading CXXYY. XX is a placeholder for the massflow, which is either 10 kg/h or 25 kg/h. YY is a placeholder for the heating current supplied to the heating element. For example C1035 refers to a configuration with a massflow of 10 kg/h and a heating current of 3.5 A.\\
Two different massflow rates were chosen to investigate the influence of turbulent mixing on the suppression of the global mode. As it turns out, the more important difference between these two configurations is a different position of the density and velocity gradient. This is discussed in \secref{sec:3}. The swirl number was adjusted to a similar value for both massflow rates. The flow is heated by forced convection from the heating element. The amount of heat transferred from the heating element to the flow is a function of the local velocity in the recirculation bubble. The flow configurations with a massflow rate of 25 kg/h thus allowed a stronger heating of the flow, with up to 6 A.   

\begin{table}
\centering
\caption{Experimental parameters.\label{tab:ExpPara}}  
\begin{ruledtabular}  
\begin{tabular}{l c c c c c c c c c c}
& $v_{bulk}$ [m/s] &  $\textrm{Re}$ & $\textrm{Ri}$ & \textrm{Ma} & \textrm{S} & $\min\left(\rho^{*}\right)$  & $\dot{m}$ [kg/h] & \pbox{2cm}{\textcolor{black}{Heating}\\\textcolor{black}{current [A]}} & \pbox{2cm}{\textcolor{black}{Heating}\\\textcolor{black}{power [W]}}\\
\hline
C1000 & 1.2 & 4000 & 0      & 0.0041 & 1.30 & 1    & 10 & 0   & -\\
C1010 & 1.2 & 4000 & 0.004  & 0.0041 & 1.30 & 0.95 & 10 & 1   & 4\\
C1020 & 1.2 & 4000 & 0.009  & 0.0041 & 1.30 & 0.80 & 10 & 2   & 16\\
C1035 & 1.2 & 4000 & 0.019  & 0.0041 & 1.30 & 0.57 & 10 & 3.5 & 52\\
\hline
C2500 & 2.9 & 10000 & 0     & 0.01 & 1.2 & 1      & 25 & 0    & -\\
C2520 & 2.9 & 10000 & 0.004 & 0.01 & 1.2 & 0.86   & 25 & 2    & 20\\
C2540 & 2.9 & 10000 & 0.014 & 0.01 & 1.2 & 0.58   & 25 & 4    & 80\\
C2560 & 2.9 & 10000 & 0.023 & 0.01 & 1.2 & 0.4    & 25 & 6    & 180\\
\end{tabular}
\end{ruledtabular}
\end{table}
%------------------------------------------------------------------------------------------------------------------------------------------------------------------------------------------------------------------
\subsection{Data acquisition}
\begin{figure}
\centering
\includegraphics{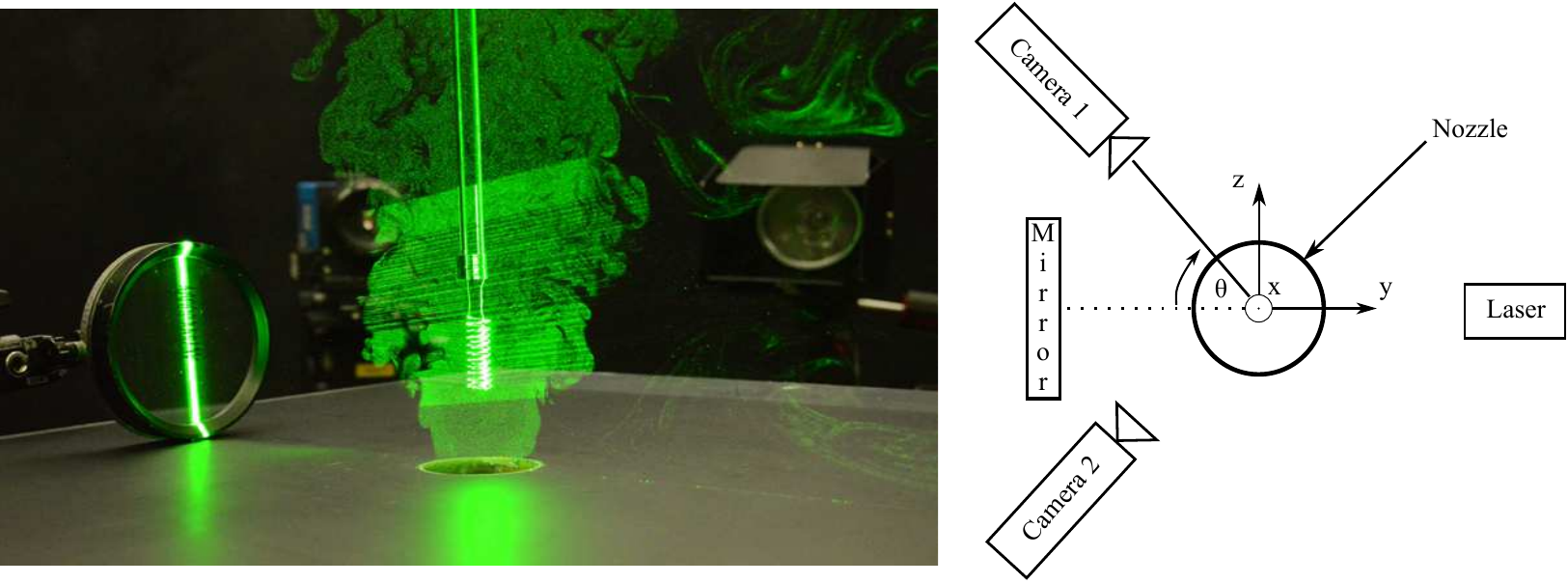}%
\caption{The photograph on the left shows the experimental setup, the laser light sheet enters from the right. The sketch on the right indicates the Stereo PIV camera arrangement.\label{fig:Camera_Setup}}%
\end{figure}

The velocity fields were measured with Stereo Particle Image Velocimetry. The system consisted of a Quantel Twins BSL 200 laser, capable of emitting 170 mJ per pulse. Images were recorded with two pco 2000 cameras with a resolution of 2048x2048 pixel. Data was acquired at a rate of 6 Hz. A total of 1000, respectively 500 snapshots were acquired for C10YY and C25YY. The photograph in \figref{fig:Camera_Setup} shows the seeded jet and the heating element. Note that an inclined mirror was used to reflect the shadow of the heating element to a position downstream of the heating element. The mirror additionally mitigated the loss of laser light intensity on the shadow side of the heating element. The camera arrangement is shown in \figref{fig:Camera_Setup}. The cameras were arranged to include an angle of 90\textdegree \;between them and each camera had an angle of 45\textdegree\; to the measurement plane. The measurement plane extended from $-2\textrm{D}$ to $2\textrm{D}$ in the $y$-direction and from 0 to $4\textrm{D}$ in the $x$-direction.\\
The double images were processed using the commercial software PIVview (PIV{\textit{TEC}} GmbH) using standard digital PIV processing \cite{Willert.1991}. The data analysis employed iterative multigrid interrogation with image deformation \cite{Scarano.2002}. The final size of the interrogation window was 32 x 32 pixel with an overlap of 50\%. Errors in the laser sheet alignment were minimized by the use of corrected mapping functions. The initial calibration datum marks were back projected onto the measurement plane by an optimized Tsai camera model \cite{Soloff.1997}\\
Temperature measurements were undertaken with a 0.25 millimeter type K thermocouple, manufactured by OMEGA Engineering. The thermocouple was connected to an in-house built amplifier circuit based on the Analog Devices AD8495 thermocouple amplifier with cold junction compensation. In order to reject noise from the temperature measurements the input of the amplifier circuit was further equipped with a first order Butterworth filter. The filter was adjusted so that a signal reduction of 3 dB is observed at 421 Hz.
%------------------------------------------------------------------------------------------------------------------------------------------------------------------------------------------------------------------
\subsection{Measurement uncertainty and convergence of mean quantities}
The measurement uncertainty of the velocity measurements is estimated in terms of the data evaluation strategy \cite{Stanislas.2008}. With a typical particle image diameter of 1 pixel Raffel et al. \cite{Raffel.1998} give an uncertainty of 0.1 pixel for the above mentioned evaluation strategy. The larger measurement uncertainty is observed for C10YY. With a pulse delay of 180 $\mu s$ and a magnification factor of seven pixel per millimeter the uncertainty is 6.7\% of the bulk velocity. It is possible to reduce this error by using larger pulse delays. However, this was not feasible in the present study, because of the large out-of-plane swirl velocity. \\
The uncertainty of the temperature measurement is influenced by the measurement accuracy of the thermocouple probe itself, the linearization of the Seebeck coefficient curve and the accuracy of the analog-digital converter. The linearization error was eliminated by using the inverse thermocouple polynomials provided in the NIST ITS-90 database \cite{Burns.1993}. Since the measurement error of the thermocouple probe is dependent on the temperature, an upper limit on the overall measurement error is calculated at the largest measured temperature for C10YY and C25YY. The largest temperatures, $\ts{T}{max}$, are measured for C1035 and C2560 and are 446\textdegree C and 217\textdegree C, respectively. The uncertainty of the National Instruments NI9215 analog-digital converter is given by the manufacturer as 0.003456V. Again referencing the largest temperatures measured, this amounts to 0.32\% and 0.15\% of the respective $\ts{T}{max}$ of C10YY and C25YY. The combined error is listed in \tabref{tab:MeanConv}. \\
To test the statistical validity of the measured mean values, we compute the standard error of the mean (SEM), as well as the upper and lower ($b_{95}^{+}, b_{95}^{-}$) bound of the 95\% confidence interval for each velocity component and the temperature. The SEM is given as
\begin{equation}
\textrm{SEM} = \frac{\sqrt{\frac{1}{N-1}\sum\limits_{i = 1}^{N}{\left(\phi_{i} - \overline{\phi}\right)^{2}}}}{\sqrt{N}}.
\end{equation}
The 95\% confidence interval is then calculated as
\begin{equation}
[b_{95}^{-}, b_{95}^{+}] = \left[\overline{\phi} - \widetilde{z}_{0.95} \textrm{SEM}, \overline{\phi} + \widetilde{z}_{0.95} \textrm{SEM} \right].\\
\end{equation}
$N$ denotes the number of samples, $\phi_{i}$ a random variable and $\overline{\phi}$ the average of that random variable. $\widetilde{z}_{0.95}$ is the value of the standard normal distribution at five percent probability of error.\\
\begin{table}
\centering
\caption{Measurement uncertainty and the 95\% confidence interval of the mean.\label{tab:MeanConv}} 
\begin{ruledtabular}    
\begin{tabular}{l c c c}
  \multicolumn{4}{c}{Measurement uncertainty velocity/temperature}   \\
\hline
C10YY & & 6.7\% $v_{bulk}$ & 1.38\% $\ts{T}{max}$ \\%Q10
C25YY & & 4.8\% $v_{bulk}$ & 1.18\% $\ts{T}{max}$ \\%Q25
\toprule
 \multicolumn{4}{c}{95\% Confidence interval mean velocity/temperature}   \\
\hline
\multirow{3}{*}{C10YY} & \ldelim\{{3}{11pt} & $b_{95}^{+}(\overline{v}_{x}) - b_{95}^{-}(\overline{v}_{x}) = 2.75\%\;v_{bulk}$ & \\
			 &		      & $b_{95}^{+}(\overline{v}_{y}) - b_{95}^{-}(\overline{v}_{y}) = 1.77\%\;v_{bulk}$ & $\left|b_{95} - \overline{T}\right| = 0.12\%\;T_{\infty}$ \\
			 &    		      & $b_{95}^{+}(\overline{v}_{z}) - b_{95}^{-}(\overline{v}_{z}) = 1.75\%\;v_{bulk}$ & \\
\multirow{3}{*}{C25YY} & \ldelim\{{3}{11pt} & $b_{95}^{+}(\overline{v}_{x}) - b_{95}^{-}(\overline{v}_{x}) = 5.15\%\;v_{bulk}$ & \\
			 &                    & $b_{95}^{+}(\overline{v}_{y}) - b_{95}^{-}(\overline{v}_{y}) = 3.22\%\;v_{bulk}$ & $\left|b_{95} - \overline{T}\right| = 0.3\%\;T_{\infty}$ \\
			 &                    & $b_{95}^{+}(\overline{v}_{z}) - b_{95}^{-}(\overline{v}_{z}) = 2.22\%\;v_{bulk}$ & \\
\end{tabular}
\end{ruledtabular}
\end{table}
%-------------------------------------------------------------------------------------------------------------------------------------------------------------------------------------------------------------------
%------------------------------------------------------------------------------------------------------------------------------------------------------------------------------------------------------------------- 
\section{Data analysis methodology}
\label{sec:2}
%------------------------------------------------------------------------------------------------------------------------------------------------------------------------------------------------------------------- 
\subsection{Triple Decomposition}
To study flows that feature a dominant coherent structure, it is convenient to adopt the triple decomposition of the flow field, introduced by Reynolds \& Hussain \cite{Reynolds.1972}. The velocity vector is decomposed into mean, coherent, and purely stochastic part,
\begin{equation}
\textbf{v}(\textbf{x},t) = \overline{\textbf{v}}(\textbf{x}) + \underbrace{\widetilde{\textbf{v}}(\textbf{x},t) + \textbf{v}^{\prime}(\textbf{x},t)}_{\widehat{\textbf{v}}},
\end{equation}
where $\widehat{\textbf{v}}$ denotes coherent and incoherent fluctuating quantities. The time average is defined as
\begin{equation}
\overline{\textbf{v}}(\textbf{x}) = \lim\limits_{a \rightarrow \infty}{\frac{1}{a}\int_{0}^{a}{\textbf{v}(\textbf{x},t) \textrm{d}t}}.
\end{equation}
The coherent velocity component is obtained by subtracting the mean flow from the phase-averaged flow
\begin{equation}
 \widetilde{\textbf{v}}(\textbf{x},t) = \langle \textbf{v}(\textbf{x},t) \rangle - \overline{\textbf{v}}(\textbf{x}),
\end{equation}
with the definition of the phase average reading
\begin{equation}
\langle \textbf{v}(\textbf{x},t) \rangle = \lim\limits_{N \rightarrow \infty}{\frac{1}{N}\sum_{0}^{N-1}{\textbf{v}(\textbf{x},t + n \tau)}},
\end{equation}
where $\tau$ denotes the period of the wave. In the present study, the data set from which these quantities are to be extracted consists of non-time-resolved uncorrelated PIV snapshots taken at arbitrary time increments. The phase-averaged velocity is reconstructed from a POD. This method is only applicable, if the limit cycle of the global mode is represented by two coupled POD modes. This has been demonstrated in the studies of Oberleithner et al. \cite{Oberleithner.2011b, Stohr.2012}, where the POD was used to extract the dominant coherent structures from swirling jet experiments.
%------------------------------------------------------------------------------------------------------------------------------------------------------------------------------------------------------------------- 
\subsection{Proper Orthogonal Decomposition}
The POD is a suitable method for the analysis of coherent structures in turbulent flow (consult Holmes et al.\cite{Holmes.1998b} for full details of the method). 
The starting point of the POD is a decomposition of the velocity vector $\textbf{v}$ into a mean $\overline{\textbf{v}}$ and fluctuating part $\widehat{\textbf{v}}$,
\begin{equation}
\textbf{v}(\textbf{x},t) = \overline{\textbf{v}}(\textbf{x}) + \widehat{\textbf{v}}(\textbf{x},t).
\end{equation}
The idea of the POD is to represent the fluctuating part as a finite series, via
\begin{equation}
\label{eqn:PODstandard}
\widehat{\textbf{v}}(\textbf{x},t) = \sum_{i=1}^{N}{a_{i}(t)\boldsymbol{\Psi}_{i}(\textbf{x})}.
\end{equation}
$\textbf{x}$ denotes the coordinate vector, $t$ the time, and $N$ the number of PIV snapshots. The decomposition is made in terms of temporal coefficients, $a_{i}(t)$, and  spatial modes, $\boldsymbol{\Psi}_{i}$. The POD provides natural sorting of the spatial modes in terms of their turbulent kinetic energy. Furthermore, the POD is optimal in the sense that there is no other truncated series expansion of a data set that has a smaller mean square truncation error \cite{Holmes.1998b}.\\
While there are a number of possibilities to compute the POD, we use the singular value decomposition to obtain the temporal coefficients, spatial modes and the singular values.  The squared singular values, $\sigma^2$, are a measure of the kinetic energy (KE) of each mode. The total KE is thus given as
$\ts{KE}{total} = \sum_{i=1}^{N}{\sigma_{i}^{2}}$.
In turbulent flows, the large scale structures of the flow usually contain a major part of the turbulent kinetic energy. Hence, the first few POD modes can be expected to provide a representation of the dominant coherent structures.\\
As in Oberleithner et al. \cite{Oberleithner.2011b}, two POD modes were identified in this study that span the basis of the global mode. From the temporal coefficients of these two modes, the phasing of the global mode can be calculated. This phase information then allows the sorting and averaging of the uncorrelated PIV snapshots according to the phase angle of the coherent structure. The coherent velocity component $\widetilde{\textbf{v}}(\textbf{x},t)$ is, hence, represented by two POD modes. 
%-------------------------------------------------------------------------------------------------------------------------------------------------------------------------------------------------------------------
\FloatBarrier
\subsection{Linear stability analysis}
We now turn to the relation between the coherent structure of the flow and linear stability analysis. 
The linear stability equations for an infinitesimal disturbance growing on a non-linearly corrected non-isothermal mean flow are given as (Reynolds \& Hussain \cite{Reynolds.1972,Lesshafft.2007})
\begin{align}
\frac{\partial \widetilde{\textbf{v}}}{\partial t} + \widetilde{\textbf{v}} \cdot \nabla \overline{\textbf{v}} &= -\frac{1}{\overline{\rho}}\nabla \widetilde{p} + \frac{1}{\mathrm{Re}}\Delta \widetilde{\textbf{v}} \label{eq:LinNav1} \\ 
\nabla \cdot \widetilde{\textbf{v}} &= 0.
\label{eq:LinNav2}
\end{align} 
$\overline{\textbf{v}}$ denotes mean flow quantities, whereas $\widetilde{\textbf{v}}$ denotes the perturbation. Small scale fluctuations can be included in the analysis via an effective viscosity $\nu = \nu_{\mathrm{molecular}} + \nu_{\mathrm{turbulent}}$ in the calculation of the Reynolds number, where $\nu_{\mathrm{turbulent}}$ is an eddy viscosity. Similar approaches have been followed by Viola et al. \cite{Viola.2014}, Oberleithner et al. \cite{Oberleithner.2011b, Oberleithner.2015c}, Marasli et al. \cite{Marasli.1994}, Reau \& Tumin \cite{Reau.2002} and Kitsios et al. \cite{Kitsios.2011}. To date no consensus has formed on how the eddy viscosity should be calculated and suggestions range from uniform eddy viscosity models \cite{Oberleithner.2011b, Oberleithner.2014}, to elaborate mixing length models \cite{Viola.2014}, to nonlinear eddy viscosity models \cite{Kitsios.2010b,Kitsios.2011} and to least square fits over all Reynolds stress components \cite{Oberleithner.2015c,Ivanova.2012}. In swirling jets undergoing vortex breakdown, the off-diagonal components of the Reynolds stress tensor are of similar magnitude in the near field. The least squares approach takes all of these into account and appears as the most suitable. We have applied it to our data and find that it has only a minute impact on the results of the stability analysis presented in \secref{sec:3}. Given that the Reynolds number is at most 10000 it is plausible that fine scale turbulence has only a minor impact on the large scale structures. We observe that the computation of the eddy viscosity and its inclusion into the analysis lends itself to numerical difficulties. Since the inclusion of an eddy viscosity does not improve the results, the analysis presented in \secref{sec:3} will rely only on the molecular viscosity.\\
For local modal stability analysis, the disturbance ansatz
\begin{equation}
\left\{\widetilde{\textbf{v}}, \widetilde{p}\right\} = \left\{iF(r), G(r), H(r), P(r)\right\}\exp(i(\alpha x + m\theta -\omega t))
\end{equation}
is substituted into \eqnref{eq:LinNav1} and \eqnref{eq:LinNav2}. $\alpha$ and $m$ denote the axial and azimuthal wave-number and $\omega$ signifies the frequency. \Eqnref{eq:LinNav2}, \eqnref{eq:LinNav2} and the disturbance ansatz together with appropriate boundary conditions pose an eigenvalue problem. The boundary conditions are given by Khorrami et al. \cite{Khorrami.1989} for mode $m$ = 1 as
\begin{equation}
F = G = H = P = 0
\end{equation}
at $r \rightarrow \infty$. Along the jet center line the boundary conditions read\\
\begin{align}
\begin{split}
&F(0) \pm G(0) = 0\\
&H(0) = P(0) = 0.
\end{split}
\end{align}
The eigenvalue problem is solved for real and complex $\alpha$, yielding a temporal, respectively spatio-temporal analysis. The first step in this analysis is to identify the appropriate mode from the temporal analysis. Each temporal stability mode is part of a Riemann sheet, the topology of which is computed by extending the analysis to complex axial wave numbers. The goal is to identify saddle points on the Riemann sheets that obey the Briggs-Bers pinch point criterion \cite{Briggs.1964}. At the location of the saddle point, the value of the absolute growth rate, $\omega_{0}$, is determined, indicating in which regions the flow is absolutely, respectively convectively unstable. The review of Huerre \& Monkewitz \cite{Huerre.1990} can be consulted for full details of the method. The location of the wavemaker is calculated from the distribution of $\omega_{0}$ via the frequency selection criterion of Chomaz et al. \cite{Chomaz.1991}. The criterion states that the frequency of the linear global mode is given by 
\begin{equation}
\omega_{\mathrm{g}} = \omega_{0}(X_s),  
\label{eq:freqSel}
\end{equation}
where $X_{s}$ is the location of a saddle point in the complex $X$-plane. This saddle point is defined by
\begin{equation}
\frac{\mathrm{d}\omega_{0}}{\mathrm{d}x}(X_s)=0.
\end{equation}
The location of the wavemaker is derived from the frequency selection criterion as $\Re(X_{s})$. Juniper \& Pier \cite{Juniper.2015} recently discussed that the region of largest structural sensitivity shows where a self-excited instability is most sensitive to changes in its internal feedback mechanism. Computing the region of structural sensitivity may provide a more consistent picture of the dynamics of coherent disturbances than the single wavemaker location derived from the frequency selection criterion. However, we find that the predictions of the frequency selection criterion are in good agreement with experimental observations and therefore use this criterion.
In order to solve the eigenvalue problem numerically, the disturbance equations are discretized using a Chebyshev pseudo spectral collocation technique \cite{Khorrami.1989} and the eigenproblem is solved employing MATLABs build in functions \texttt{eig} and \texttt{eigs}. \\
\textcolor{black}{The results displayed in \secref{sec:3} show that the flow under investigation is non-parallel. What results can be expected from a local stability analysis of this flow?\\
For the local analysis to be strictly valid, the quotient $\epsilon$, which defines the ratio of the axial instability wave length, $\lambda$, and a characteristic length scale for the development of the flow, L, has to be small. In our study $\lambda$ is typically around 45mm and L is 65mm. Following Juniper et al. \cite{Juniper.2011} we have taken the distance between the nozzle exit and the downstream end of the recirculation bubble as the flow length scale. Under these conditions $\epsilon$ has a value of around 0.7 which is clearly not negligible and can impact the accuracy of the analysis \cite{Juniper.2011,Qadri.2015}. However, Juniper et al. \cite{Juniper.2011} also point out that the accuracy of the local predictions depends on the degree of non-parallelity in the vicinity of the wavemaker and thus needs to be assessed on a case to case basis. As we point out in the following, a global stability analysis is not a viable alternative for the present data set and therefore a comparison to measured data is the best benchmark available.\\
The studies of Juniper et al. and Leontini et al. \cite{Juniper.2011,LEONTINI.2010} conduct global stability analyses that take the non-parallel effects into account and show that these deliver a more accurate prediction of the global modes growth rate and frequency. However, the elliptic operators of global stability analysis are extremely sensitive to boundary conditions and unsuitable boundary conditions can render a global analysis even more inaccurate than a local analysis \cite{Paredes.2016}. Paredes et al. \cite{Paredes.2016} highlight the severity of this problem in the context of experimentally obtained data, where vortex breakdown occurs close to the nozzle and no data upstream of the nozzle is available.\\
A number of studies assessed the ability of local linear stability analysis to identify the region of a flow that drives the (absolute) global instability \cite{GIANNETTI.2007,Juniper.2011,Juniper.2015}. These studies found that the position of the wavemaker coincides very well with regions of largest structural sensitivity derived from global stability analyses. Particularly, the study of Qadri et al. \cite{Qadri.2015} underlines the ability of local stability analysis to correctly identify the mechanisms that are responsible for different global modes, even though the flow is non-parallel.}
%------------------------------------------------------------------------------------------------------------------------------------------------------------------------------------------------------------------- 
%------------------------------------------------------------------------------------------------------------------------------------------------------------------------------------------------------------------- 
\section{Results}
\label{sec:3}
This section presents the experimental and theoretical results. The influence of the heating element on the flow field is discussed in \appref{sec:HE}. The mean velocity and temperature fields are presented first, followed by an analysis of the properties of the global mode at various heating levels. The results of the stability analysis applied to the measured data and a parametric study are outlined in the remainder of this section.
%------------------------------------------------------------------------------------------------------------------------------------------------------------------------------------------------------------------- 
\subsection{Experimental findings}
%-------------------------------------------------------------------------------------------------------------------------------------------------------------------------------------------------------------------
\subsubsection{Mean Velocity and Temperature fields}
\label{sec:MVaT}
\begin{figure}
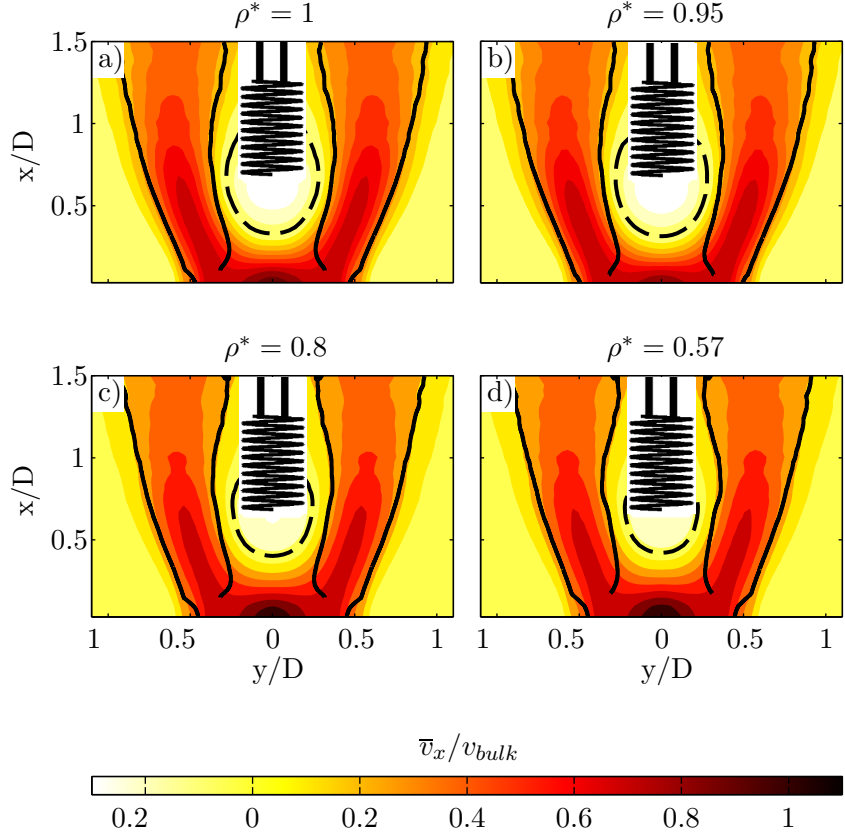
\caption{Time-averaged axial velocity of C10YY. The dashed line indicates the contour of zero axial velocity. The solid lines indicate the position of the outer and inner shear layer, respectively.\label{fig:vMeanQ10}}%
\end{figure}

\begin{figure}
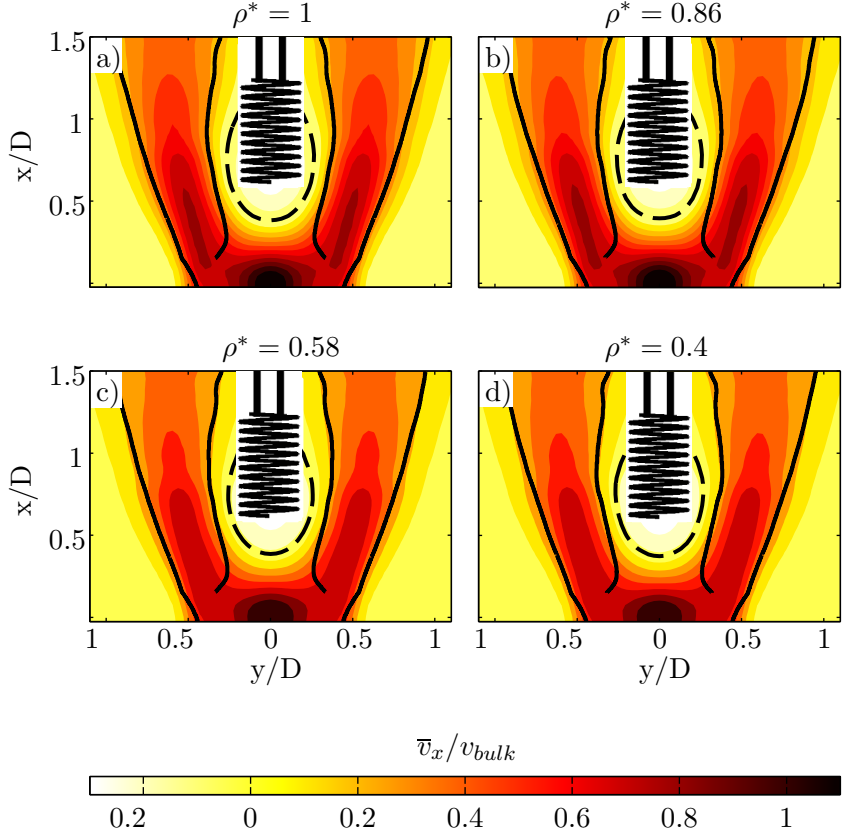
\caption{Time-averaged axial velocity of C25YY. The dashed line indicates the contour of zero axial velocity. The solid lines indicate the position of the outer and inner shear layer, respectively.\label{fig:vMeanQ25}}%
\end{figure}

\begin{figure}
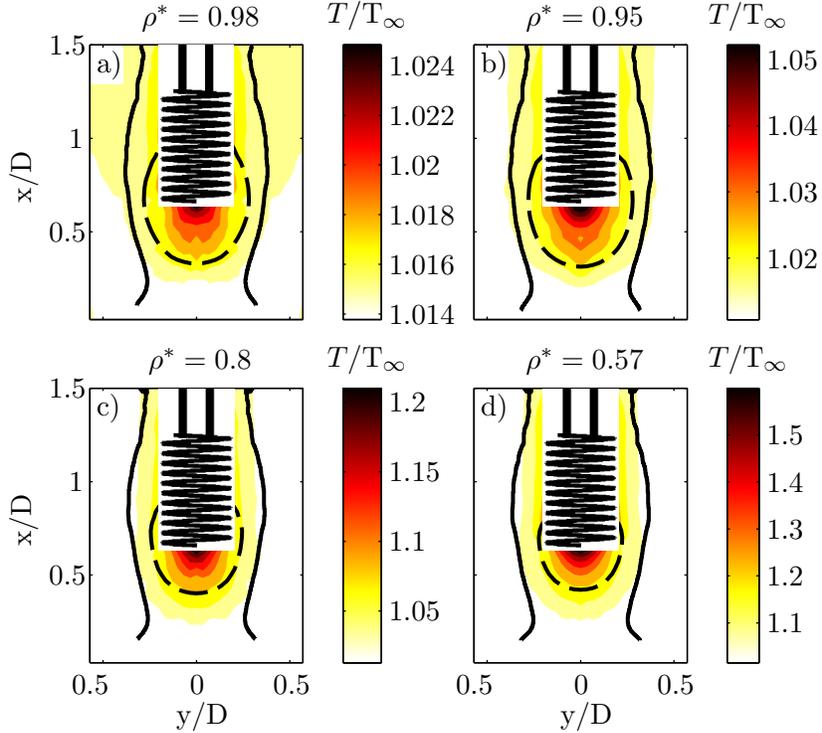
\caption{Time-averaged temperature field of C10YY. Note the different temperature scales, indicating the heating. The dashed line indicates the contour of zero axial velocity. The solid line shows the position of the inner shear layer. \label{fig:TMeanQ10}}%
\end{figure}

\begin{figure}
\includegraphics{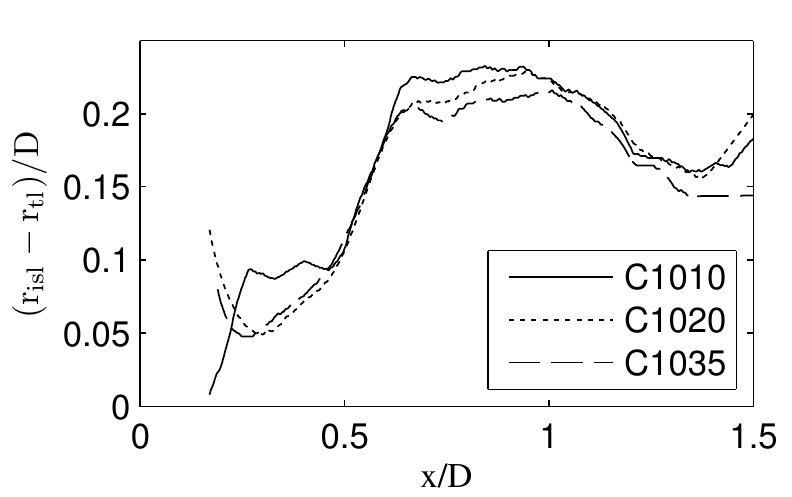}%
\caption{Radial distance between the inner shear layer and the temperature mixing layer versus streamwise distance for C1010, C1020 and C1035. \label{fig:VeloTMeas}}%
\end{figure}

\Figref{fig:vMeanQ10} a) shows the time-averaged axial velocity of C1000 at isothermal conditions, whereas \figref{fig:vMeanQ10} b) - d) present the axial velocity at different heating. All velocity fields exhibit the typical pattern of a swirling jet undergoing vortex breakdown. The jet emanating from the nozzle is forced into an expansion around the recirculation bubble. The contour line of zero axial velocity is indicated by the thick blue line in \figref{fig:vMeanQ10} a) - d). From \figref{fig:vMeanQ10} a) to \figref{fig:vMeanQ10} b) the size of the recirculation bubble grows slightly. While the lateral extent remains unaltered, the upstream stagnation point shifts from 0.33D in \figref{fig:vMeanQ10} a) to 0.31D in \figref{fig:vMeanQ10} b). The extent of the recirculation decreases in \figref{fig:vMeanQ10} c) and d).\\ 
The time-averaged axial velocity of C25YY is shown in \figref{fig:vMeanQ25}. \figref{fig:vMeanQ25} a) again presents the velocity field at isothermal conditions, whereas \figref{fig:vMeanQ25} b) - d) are the velocity fields at larger temperatures. Although C10YY and C25YY seem qualitatively similar, C25YY shows a very different response to the heating. It shows no alteration of the recirculation bubble, although the largest temperature investigated in C2560 is more than twice the largest temperature in C1035.\\ 
The time-averaged temperature field is exemplified with C10YY and is displayed in \figref{fig:TMeanQ10}. \figref{fig:TMeanQ10} a) - d) reveal that the temperature distribution remains similar for all heating regimes. The difference between \figref{fig:TMeanQ10} a) - d) is the maximum temperature in the vicinity of the upstream end of the heating element. In all cases, the region with the largest temperatures is confined within the recirculation bubble, which is indicated by the dashed line. The solid line shows the position of the inner-shear layer. The radial temperature gradient is mostly present in the inner shear layer.\\
To identify the position of the shear layer, two definitions for the inner and outer shear layer are used, respectively. The definition of the shear layer position, reading $\ts{r}{osl}(x) = r(\overline{v}_{x}/\overline{v}_{x\mathrm{max}} = 2)$, is used to identify the position of the outer shear layer. This criterion is not best suited for a description of the inner shear layer, because the inner shear layer forms as the axial velocity overshoot on the jet center line develops into a velocity deficit. To identify the position of the inner shear layer, we first define the difference between the local axial velocity maximum and minimum,
$\Delta(x) = \max\limits_r\left(\overline{v}_{x}(r,x))\right) - \min\limits_{r^{*}}\left(\overline{v}_{x}(r^{*},x)\right)$,
with $0<r^{*}<r_{\mathrm{max}}$. The position of the inner shear layer is identified as the radial position, at which the axial velocity is equal to half of the velocity span between the velocity maximum and minimum, reading $\ts{r}{isl}(x) = r(v_{x} \equiv \min\limits_{r^{*}}\left(v_{x}(r^{*},x)+\frac{\Delta}{2}\right))$.\\
For the non-isothermal flow, the radial position of the temperature mixing layer is likewise important. To extract the position of the mixing layer from our experimental data, we define a quantity similar to the position of the outer shear layer. This definition reads $\ts{r}{tl}(x) = r((T(r,x)-\mathrm{T}_{\mathrm{min}}(x))\equiv(\mathrm{T}_{\mathrm{max}}(x)-\mathrm{T}_{\mathrm{min}}(x))/2)$. \Figref{fig:VeloTMeas} shows the difference between the inner shear layer position and the position of the mixing layer for C10YY. The difference between the position of the shear and mixing layer is a measure of the collocation of the velocity and temperature gradient. All curves shown in \figref{fig:VeloTMeas} show similar characteristics downstream of 0.5 $x$/D.  The curves of C1020 and C1035 also show a similar trend upstream of that position. C1010 shows a distinctively different behavior. At the upstream end of the inner shear layer the positions of the mixing layer and the shear layer coincide, indicating that the velocity and temperature gradient are collocated. 
%-------------------------------------------------------------------------------------------------------------------------------------------------------------------------------------------------------------------
\FloatBarrier
\subsubsection{Impact of heating on the global mode limit cycle}
\begin{figure}
\includegraphics{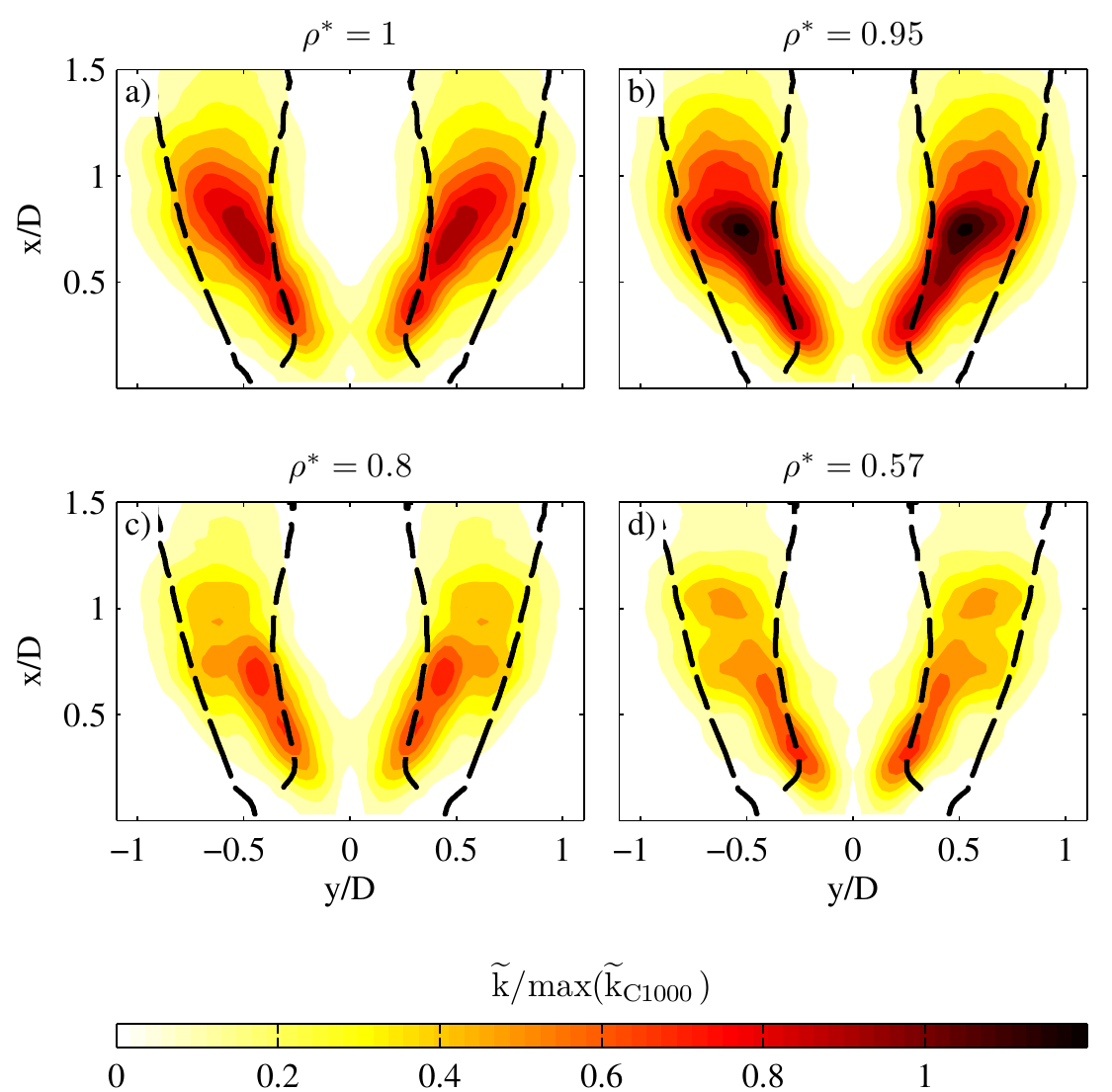}%
\caption{Evolution of the coherent kinetic energy of the global mode of C10YY as heating is applied. Dashed lines indicate the position of the inner and outer shear layer.  \label{fig:KoKiEQ10}}%
\end{figure}

\begin{figure}
\includegraphics{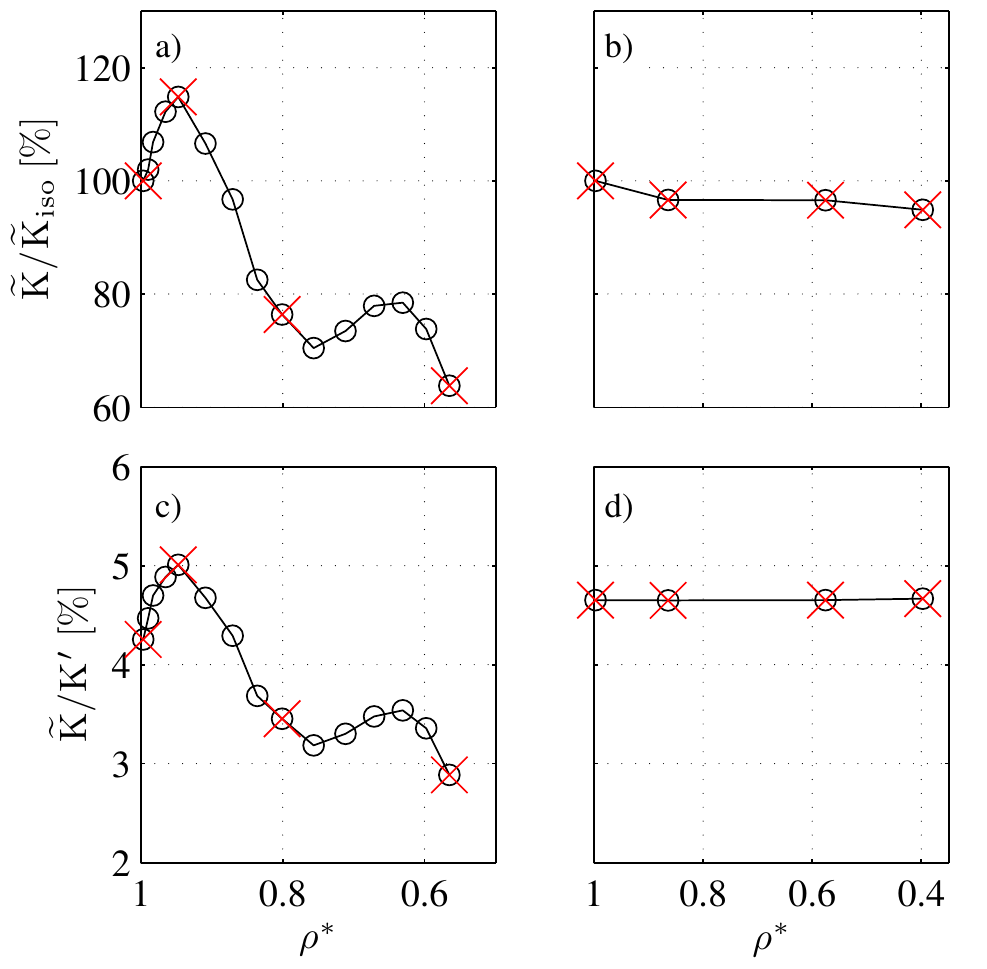}%
\caption{\textcolor{black}{a) and b): Evolution of the coherent kinetic energy $\widetilde{\mathrm{K}}$ in relation to the coherent kinetic energy at isothermal conditions $\ts{\widetilde{K}}{iso}$ as heating is applied for C10YY and C25YY, respectively. The heating increases from isothermal conditions with a density ratio of $\rho^{*} = 1$ to smaller values of $\rho^{*}$. c) and d): Relative contribution of the coherent kinetic energy $\widetilde{\mathrm{K}}$ to the total turbulent kinetic energy, $\mathrm{K}'$, for C10YY and C25YY, respectively. The measurements marked by red crosses correspond to C10YY and C25YY. Circles indicate additional configurations, for which PIV measurements were conducted.}    \label{fig:KoKiEint}}%
\end{figure}

\begin{figure}
\includegraphics{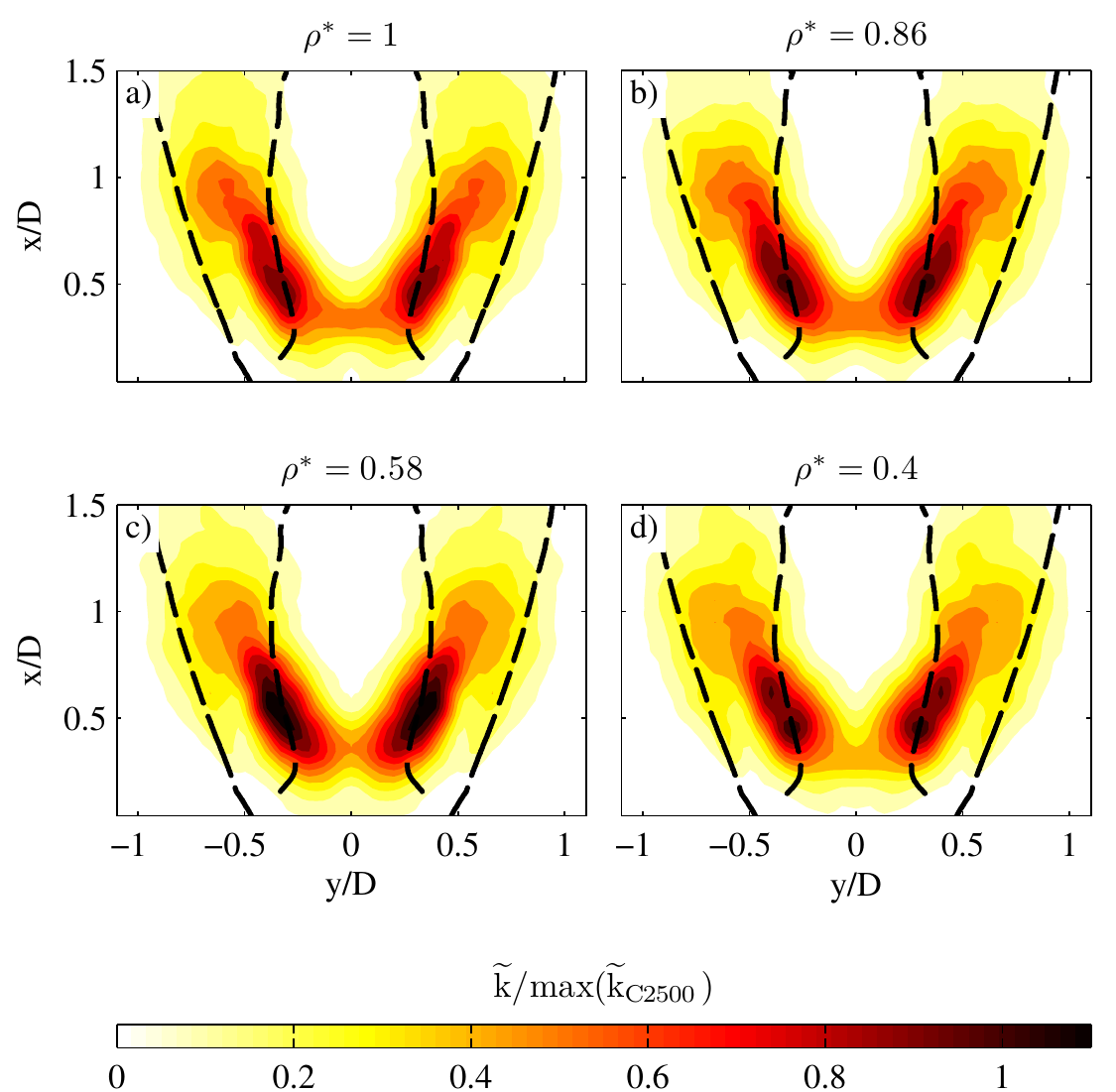}%
\caption{Evolution of the coherent kinetic energy of the global mode of C25YY as heating is applied. Dashed lines indicate the position of the inner and outer shear layer.\label{fig:KoKiEQ25}}%
\end{figure}

The impact of heating on the coherent structures associated with the global mode is discussed in this section. The coherent velocity was extracted from a POD analysis of the PIV data. The symmetry properties of the helical global mode lead us to consider only the anti-symmetric part (with respect to a cartesian coordinate system) of the velocity fluctuations for the POD analysis. The symmetry properties of the global mode and its identification from a POD analysis are thoroughly discussed in Oberleithner et al. \cite{Oberleithner.2011b}. From the two POD modes that represent the global mode, the coherent kinetic energy is calculated via
\begin{align}
\widetilde{\mathrm{k}} &= \widetilde{v}_{x}^{2} + \widetilde{v}_{y}^{2} + \widetilde{v}_{z}^{2}\\
											 &= \left|\widetilde{\textbf{v}}\right|^{2}.
\end{align}
A similar definition holds for the mean kinetic energy, $\overline{\mathrm{k}}$, and the kinetic energy of the incoherent fluctuations, $\mathrm{k}^{'}$. \\
We begin with the discussion of C10YY. \Figref{fig:KoKiEQ10} presents the coherent kinetic energy of the global mode for C1000, C1010, C1020 and C1035. The position of the inner and outer shear layer is given for reference by the dashed lines. At isothermal conditions, \figref{fig:KoKiEQ10} a), the coherent kinetic energy is mainly distributed in the inner shear layer. The overall distribution of energy remains similar for all heating levels presented in \figref{fig:KoKiEQ10} a) - d). From nearly isothermal conditions \figref{fig:KoKiEQ10} a) to an intermediate heating regime with $T_{max}/\mathrm{T}_{\infty} = 1.71$ at \figref{fig:KoKiEQ10} b), an increase in energy is evident. After the increase, the modal energy is strongly reduced for stronger heating, \figref{fig:KoKiEQ10} c) and d). The difference in modal energy is quantified with the relative energy content of the global mode. This quantity is obtained from the POD analysis, via
\begin{equation}
\frac{\widetilde{\mathrm{K}}}{\mathrm{K}^{'}} = \frac{\int_{0}^{\infty}{\int_{0}^{\infty}{\widetilde{\mathrm{k}}\,r\,\mathrm{d}r\mathrm{d}x}}}
{\int_{0}^{\infty}{\int_{0}^{\infty}{\mathrm{k}^{'}\,r\,\mathrm{d}r\mathrm{d}x}}}.
\end{equation}
To assess only the change in energy content of the global mode, relative to isothermal conditions, we compute
$\widetilde{\mathrm{K}}/\widetilde{\mathrm{K}}_{\mathrm{iso}}$
for the respective heating configuration. The evolution of $\widetilde{\mathrm{K}}/\widetilde{\mathrm{K}}_{\mathrm{iso}}$ is presented in \figref{fig:KoKiEint} a) and b) for C10YY and C25YY, respectively. The values of $\widetilde{\mathrm{K}}/\mathrm{K}^{'}$ is indicated in \figref{fig:KoKiEQ10} c) for C10YY and in \figref{fig:KoKiEQ10} d) for C25YY.
 \Figref{fig:KoKiEint} a) confirms the qualitative observations made from \figref{fig:KoKiEQ10}. At a slight temperature increase, C1010, the energy of the global mode increases by 18.6\% compared to isothermal conditions. A steep decrease in the energy of the global mode can be observed from C1010 to C1020. Compared to the energy content at C1010 the energy of the global mode is decreased by almost 50\%. After the steep decrease in modal energy only a slight reduction in modal energy is evident from \figref{fig:KoKiEint} a). For larger temperatures, the reduction in energy flattens, with only an additional 5\% decrease from C1020 to C1035.\\
We turn now to C25YY. From \figref{fig:KoKiEQ25} it is evident that heating has a much smaller effect on the distribution of coherent kinetic energy compared to C10YY. At all heating conditions the coherent kinetic energy remains amassed in the inner shear layer with no evident change in the absolute value of energy. This is confirmed by the values of $\widetilde{\mathrm{K}}/\widetilde{\mathrm{K}}_{\mathrm{iso}}$ presented in \figref{fig:KoKiEint}. Despite the overall larger temperatures attained in C25YY only a very small change in the global mode is evident. 

%-------------------------------------------------------------------------------------------------------------------------------------------------------------------------------------------------------------------
\FloatBarrier
\subsection{Stability Analysis of Experimental Data}
%-------------------------------------------------------------------------------------------------------------------------------------------------------------------------------------------------------------------
%-------------------------------------------------------------------------------------------------------------------------------------------------------------------------------------------------------------------
\subsubsection{Experimental Input}
\label{sec:EIn}
We turn now to the analysis of the velocity fields of C10YY and C25YY. The influence of the temperature variation is included in the analysis via the density. Assuming an ideal gas with constant mean pressure throughout the flow field, the non-dimensional equation of state reduces to \cite{Yu.1990}
\begin{equation}
\overline{\rho}(r) = \frac{1}{\overline{T}(r)}.
\end{equation}
The dependence of the dynamic viscosity on the temperature is taken into account via Sutherland's formula. With the far-field viscosity $\mu^{*}_{\infty}$ as reference, it takes the non-dimensional form
\begin{align}
\mu(T) &= T^{3/2}\left(\frac{1+\mathrm{C}_{\infty}}{T+\mathrm{C}_{\infty}}\right),\\
\intertext{with}\\
\mathrm{C}_{\infty} &= 120 K/T^{*}_{\infty}.
\end{align}
All velocities are made non-dimensional with the local velocity maximum at each flow slice. We point out that no temperature data is available from $x > 0.6 \mathrm{D}$ in the range $-0.1 < y < 0.1$. The density data in this region was interpolated from the surrounding data.
\subsubsection{The region of absolute instability of mode m = 1}
\begin{figure}
\includegraphics{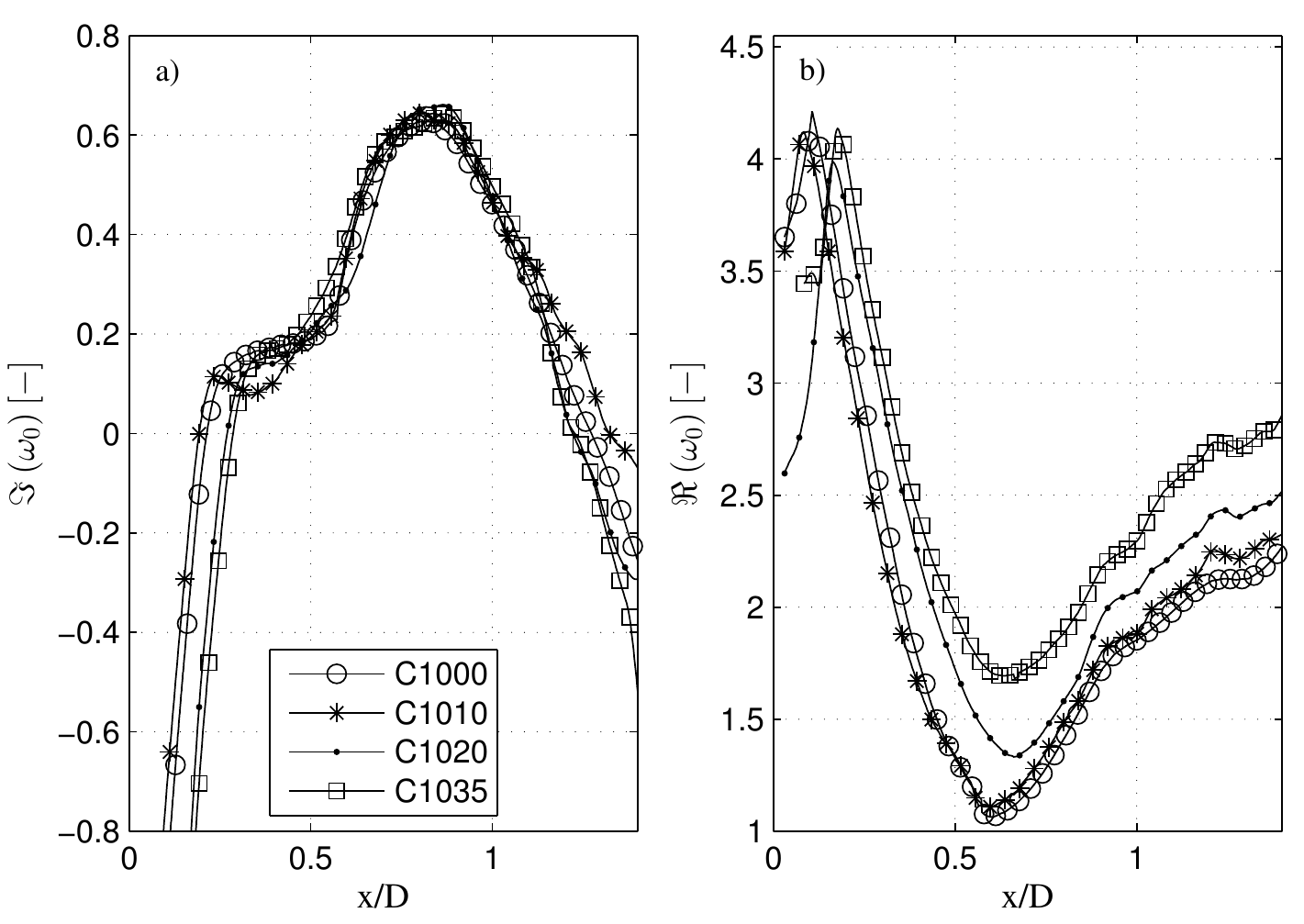}%
\caption{The absolute growth rate of mode $m$ = 1 for C10YY is shown in a). \textcolor{black}{The corresponding absolute frequency is indicated in b). Note that the legend in a) also applies to b).}\label{fig:Absamp10}}%
\end{figure}

\begin{figure}
\includegraphics{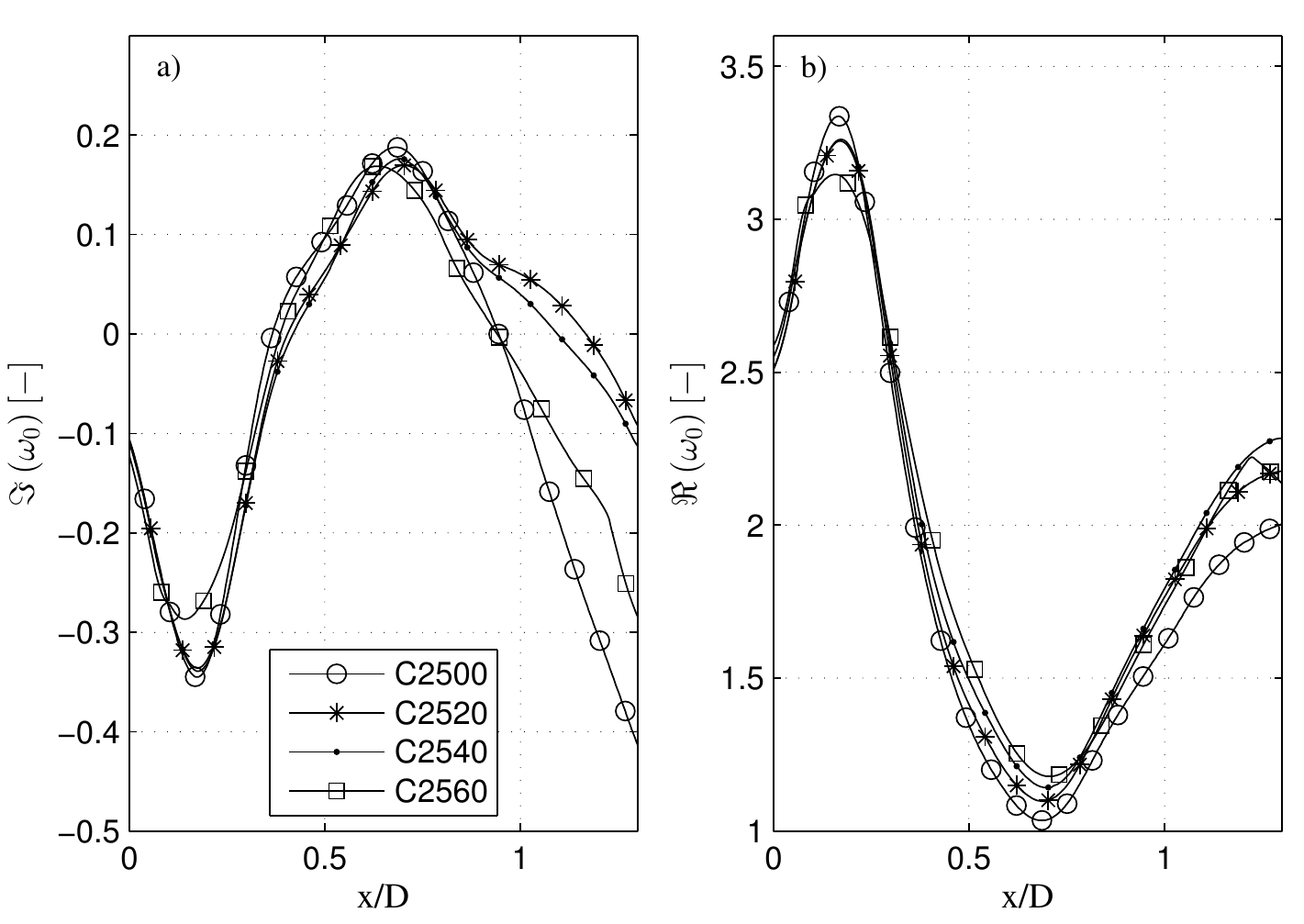}%
\caption{The absolute growth rate of mode $m$ = 1 for C25YY is shown in a). \textcolor{black}{The corresponding absolute frequency is indicated in b). Note that the legend in a) also applies to b).}\label{fig:Absamp25}}%
\end{figure}

We begin our discussion with C10YY. \Figref{fig:Absamp10} shows the absolute growth rate of the different configurations of C10YY. The largest absolute growth rates remain the same to a large extent for all configurations considered. Differences prevail in the size of the region of absolute instability and in the upstream axial location, where the convective/absolute instability transition takes place. From isothermal conditions, C1000, to slight heating at C1010, the axial extent of the region of absolute instability increases by 7\%. The convective/absolute instability transition is at the most upstream position of all configurations. At a smaller density ratio, C1020, the extent of the region of absolute instability has decreased by 11\%, relative to isothermal conditions. Note that the upstream end of the region of absolute instability has shifted downstream compared to isothermal conditions. At the smallest density ratio, C1035, the region of absolute instability is 13\% smaller than at isothermal conditions. For the strongest heating, convective-absolute transition occurs at the most downstream position. The axial extent of the region of absolute instability appears to be correlated to the change in modal energy computed from the POD. For C1010 an increase in both quantities can be observed, whereas the energy content and the region of absolute instability are reduced for C1020 and C1035. While we find a correspondence between the size of the region of absolute instability and the energy content of the global mode observed in the experiments, no direct correlation between the largest value of the absolute growth rate and the suppression of the global mode is evident from \figref{fig:Absamp10}. \\  
Considering the higher Reynolds number case, \figref{fig:Absamp25}, as the density ratio decreases from C2500 through C2560, the axial extent of the region of absolute instability first increases by 33\% and 13\% and than decreases by 4\%, relative to isothermal conditions. Note that the growth in the region of absolute instability is mostly evident for $x/\mathrm{D} > 0.8$, where the PIV data quality degraded due to the shadow of the heating element. Upstream of the heating element at $x/\mathrm{D} = 0.6$ only minor changes in the region of absolute instability occur. This is consistent with the experimental observations, where for example \figref{fig:KoKiEint} b) indicates no substantial change in the strength of the global mode. \\
Although the region of absolute instability seems to be a good indicator for global instability, it does not explain the different response of the two Reynolds number cases to the heating. To investigate this question further, the predictions of the frequency selection are addressed next. 
%---------------------------------------------------------------------------------------------------------------
\FloatBarrier
\subsubsection{The wavemaker of mode m = 1}
\begin{figure}
\includegraphics{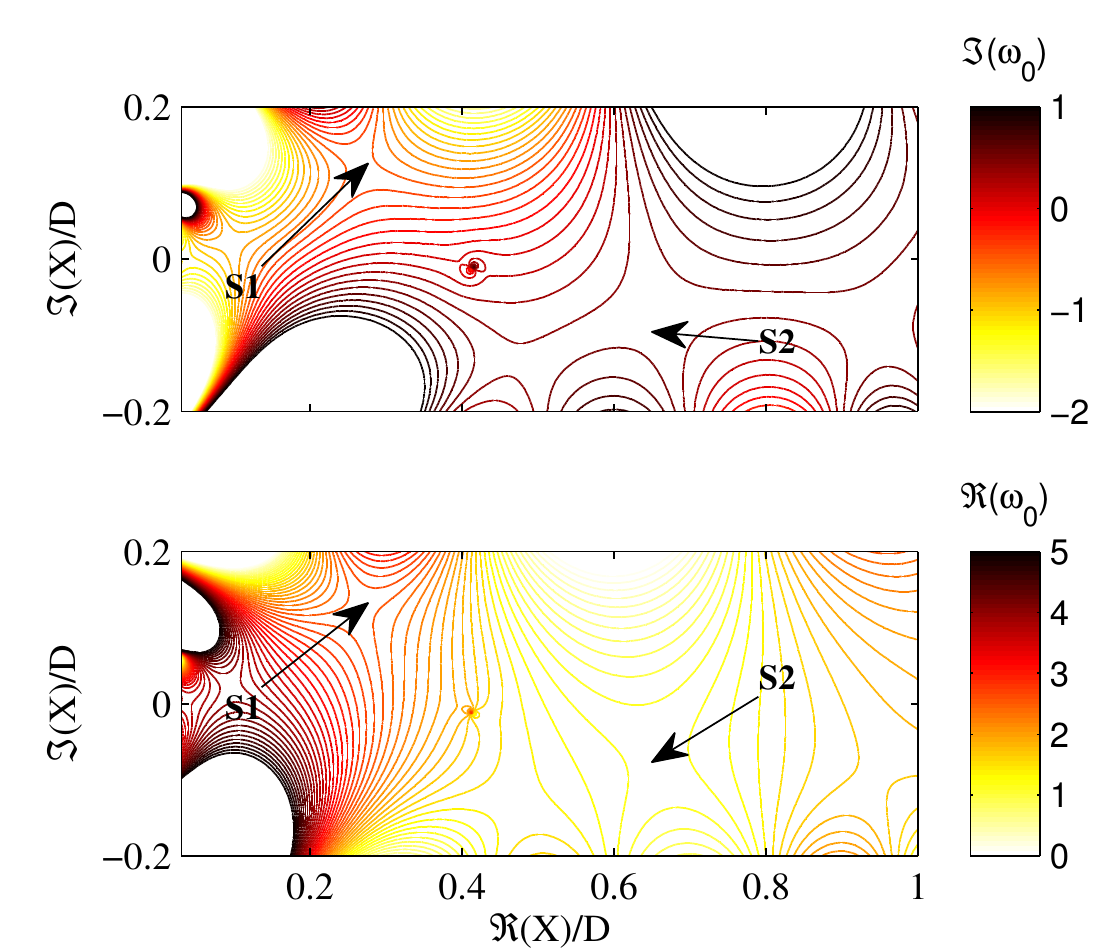}%
\caption{\textcolor{black}{a): Real part of the extension of the absolute growth rate curve into the complex $X$-plane for C1035. b): The corresponding imaginary part. The selected saddle point is marked by an \textbf{S}.}\label{fig:freqSelQ10}}%
\end{figure}

\begin{figure}
\includegraphics{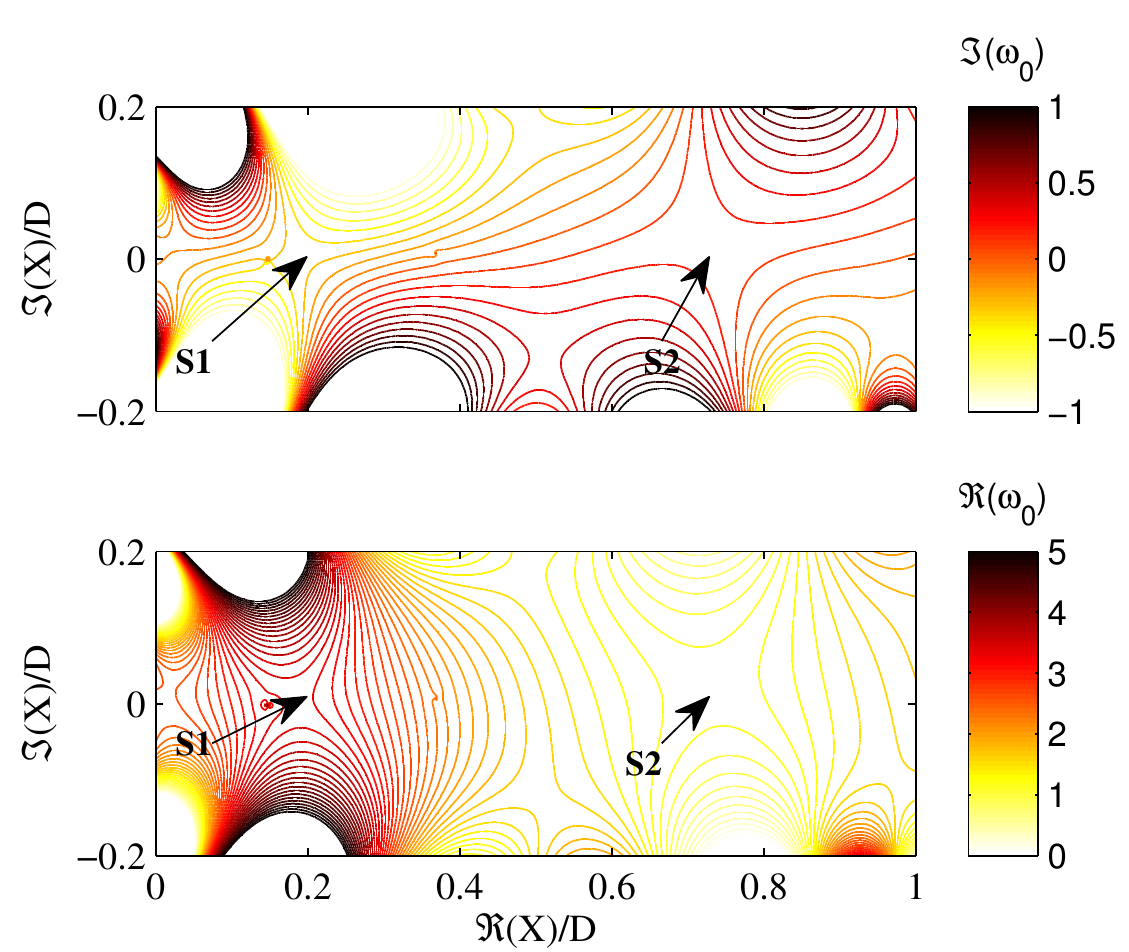}%
\caption{\textcolor{black}{a): Real part of the extension of the absolute growth rate curve into the complex $X$-plane for C2560. b): The corresponding imaginary part. The selected saddle point is marked by an \textbf{S}.}\label{fig:freqSelQ25}}%
\end{figure}

\begin{figure}
\includegraphics{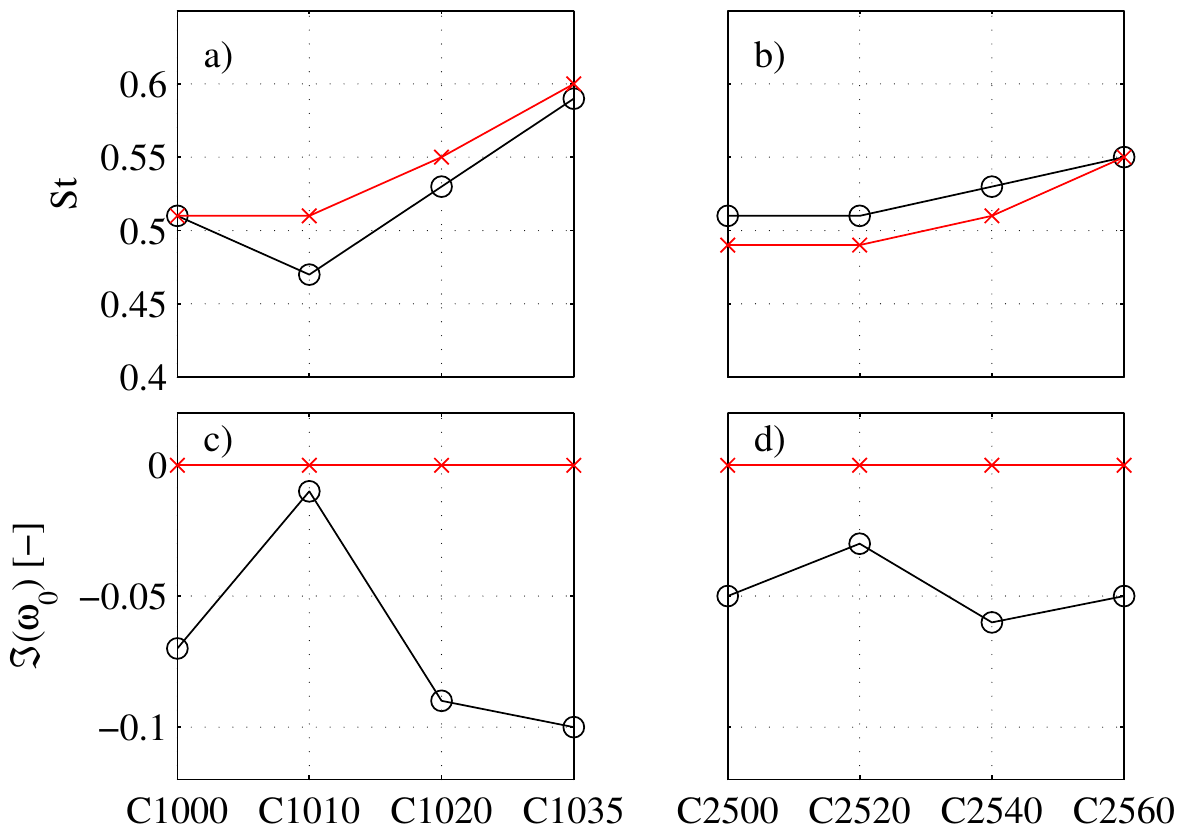}%
\caption{Strouhal number of the global mode computed from the frequency selection criterion (black circles) and from hotwire measurements (red crosses) for C10YY (a)) and C25YY (b)). The growth rate of the global mode derived from the frequency selection criterion is indicated by black circles in c) for C10YY and in d) for C25YY, respectively. Red crosses mark the global mode growth rate that is expected on the limit cycle. \label{fig:freqSelPlot}}%
\end{figure}

In order to identify the wavemaker location for each configuration, the frequency selection criterion of Chomaz et al. \cite{Chomaz.1991} is applied to the absolute growth rate curves presented in \figref{fig:Absamp10} and \figref{fig:Absamp25}. The analytical continuation into the complex plane is computed from a Pad\'{e} approximant to each of the absolute growth rate curves \cite{Juniper.2011}. The Pad\'{e} approximant is of the form 
\begin{equation}
f(x) = \frac{P(x)}{Q(x)} = \frac{a_{0} + a_{1}x + \cdots + a_{n}x^{n}}{1 + b_{1}x + \cdots + b_{m}x^{m}}.
\end{equation}
Pad\'{e} approximants can provide a very good fit to the data at low polynomial order, minimizing the number of spurious saddles in the complex X-plane. Eighth to tenth order polynomials were used.
%\Secref{sec:fitQuali} shows the quality if the fit. 
\Figref{fig:freqSelQ10} and \figref{fig:freqSelQ25} show the saddle points in the complex X-plane that define the wavemaker location for the two configurations with the smallest density ratio, C1035 and C2560. We find a similar topology for each configuration of C10YY and C25YY. The frequency and the growth rate of the global mode are determined at the saddle points \textbf{S1} shown in \figref{fig:freqSelQ10} and \figref{fig:freqSelQ25}.\\
The dimensionless frequency is presented as a Strouhal number, St. The Strouhal number is based on the respective bulk velocity and the nozzle diameter, $\mathrm{St} = fD/v_{bulk}$. \Figref{fig:freqSelPlot} a) and b) show the computed Strouhal numbers as black circles, together with the value of the Strouhal number derived from the hotwire measurements for C10YY and C25YY, respectively. A good agreement between the predicted and measured frequency is evident. The largest deviation of the predicted Strouhal number from the measured Strouhal number is 8\% and is observed for C1010. The growth rate predicted from the frequency selection criterion is presented in \figref{fig:freqSelPlot} c) and d) for the configurations C10YY and C25YY. The red crosses indicate that the growth rate of the global mode should be ideally equal to zero on the limit cycle \cite{Barkley.2006,Oberleithner.2011b,Liang.2005}. The predicted growth rate is negative for all configurations considered, but with an absolute value close to zero. The deviation from zero may be a sign of the non-parallelism of the underlying mean-flow. Bearing in mind that local linear stability analysis is applied here to a strongly non-parallel flow, the agreement of the predicted Strouhal numbers with the experimental results are astonishing and clearly show the physical relevance of the frequency selection criterion. 
The small deviation from a growth rate of zero indicates that the nonlinear saturation of the global mode is mainly manifested by a mean flow correction and nonlinear mode-mode interactions are not relevant for this flow.\\
We continue the analysis with the wavemaker location, which is tabulated in \tabref{tab:freqSel}, together with the density ratio at the wavemaker location. The wavemaker for C25YY is located in regions with no density variation or only a very slight deviation from isothermal conditions. The position of frequency selection for C10YY is located at a similar axial station as for C25YY. The density ratio at the wavemaker location deviates significantly from one only for C1020 and C1035, where the global mode was observed to be suppressed in the experiments. The overall larger density ratio of C25YY is not reflected in the density ratio at the wavemaker location. The decisive contribution to the suppression of the global mode is thus not the overall density ratio, but the density at the wavemaker location.
 
\begin{table}
\centering
\caption{The wavemaker location, $\ts{x}{WM}$, together with the density ratio at the wavemaker location, $\rho^{*}@\ts{x}{WM}$, for C10YY and C25YY, respectively. \label{tab:freqSel}}  
\begin{ruledtabular}  
\begin{tabular}{c c c c c c}
%& \multicolumn{4}{c}{$C1$} & \multicolumn{4}{c}{$C2$}\\
%\hline
 %& St & $\Im(\omega_{\mathrm{g}})$ &$\ts{x}{S}/\mathrm{D}$ & $\rho^{*}@\ts{x}{WM}$&  & St & $\Im(\omega_{\mathrm{g}})$ & $\ts{x}{S}/\mathrm{D}$ & $\rho^{*}@\ts{x}{WM}$\\
%\hline
 %C1000   & 0.51 (0.51) & -0.07i & 0.13 & 1  & C2500 & 0.51 (0.49) & -0.05i & 0.15 &1  \\
 %C1010   & 0.53 (0.51) & +0.38i & 0.08 & 0.97 & C2520& 0.53 (-)     & -0.06i  & 0.18 &0.94   \\
 %C1020   & 0.54 (0.55) & -0.12i  & 0.17 & 0.78 & C2540 & 0.55 (-)    & -0.08i  & 0.21 &0.92   \\
 %C1035   & 0.59  (0.6) & -0.1i & 0.19 & 0.68 & C2560 & 0.55 (0.55)& -0.05i  & 0.19 &0.94  \\

 &$\ts{x}{WM}/\mathrm{D}$ & $\rho^{*}@\ts{x}{WM}$&  &  $\ts{x}{WM}/\mathrm{D}$ & $\rho^{*}@\ts{x}{WM}$\\
\hline
 C1000   & 0.13 & 1    & C2500 & 0.15 &1  \\
 C1010   & 0.22 & 0.97 & C2520 & 0.18 &0.99   \\
 C1020   & 0.25 & 0.92 & C2540 & 0.20 &0.98   \\
 C1035   & 0.27 & 0.88 & C2560 & 0.19 &0.97  \\
\end{tabular}
\end{ruledtabular}
\end{table} 

\textcolor{black}{We complement this section by a discussion of the saddle point selection in the complex X-plane. Figures \figref{fig:freqSelQ10} and \figref{fig:freqSelQ25} show multiple saddle points and the question arises which one corresponds to the global mode wavemaker. Within the framework of linear stability analysis of weakly nonparallel flows, the global mode is determined by the dominant saddle point with the largest growthrate $\Im(\omega_0)$.  It is are marked as \textbf{S2} in the figs.~\figref{fig:freqSelQ10} and \figref{fig:freqSelQ25}. Their corresponding locations, frequencies and growthrates are given in table \ref{tab:freqSel2} together with the values at the subdominant saddle point \textbf{S1}.  
The wavemnaker location corresponding to the dominant saddle point is located in the center of the recirculation zone, with a frequency much lower than the measured limit-cycle frequency. 
The corresponding growthrates are slightly above zero indicating a globally unstable mean flow.} 
\begin{table}
\textcolor{black}{
\centering
\caption{The wavemaker location, $\ts{x}{WM}$, the Strouhal number and growth rate of the global mode at \textbf{S1} and \textbf{S2} for C1035 and C2560, respectively. \label{tab:freqSel2}}  
\begin{ruledtabular}  
\begin{tabular}{c c c c}
                       &$\ts{x}{WM}/\mathrm{D}$ & St [-] &  $\Im(\omega_{0})$\\
\hline
 \textbf{S1} @ C1035   & 0.27                   & 0.59   & -0.1 \\
 \textbf{S2} @ C1035   & 0.73                   & 0.25   & 0.1 \\
 \textbf{S1} @ C2560   & 0.19                   & 0.55   & -0.05 \\
 \textbf{S2} @ C2560   & 0.66                   & 0.16   & 0.02 \\
\end{tabular}
\end{ruledtabular}
}
\end{table} 

\textcolor{black}{
In consideration of the importance of the wavemaker location for the argumentation in this study, we list here our arguments why to choose  the saddle point \textbf{S1} as the relevant one. 
\begin{itemize}
	\item \textbf{Best match with measured frequency:} The subdominant saddle point \textbf{S1} predicts the measured limit-cycle frequency with acccuracy of less than $8$~\%, while the frequency at the dominant saddle point \textbf{S2} deviates by more than $50$~\%. 
	\item \textbf{General inaccuracies in growthrate prediction:} The difference in growthrate between the dominant and the subdominant saddle point is in the same range as the discrepancy between measured and predicted frequencies. Hence, the ranking between dominant and subddominant saddle points is  subjected to uncertainty. Potential sources for inaccuracies are nonparallelity of the base flow, negligence of turbulent fluctuations, and measurement uncertainty near the heating element.  
	\item \textbf{Agreement with literature:} There are several examples in literature that indicate that the wavemaker is located at or slightly upstream of the first stagnation point and not in the center of the recirculation bubble. Liang \& Maxworthy conducted time-resolved PIV measurements of a turbulent jet undergoing a swirl transient \cite{Liang.2005}. They recorded the evolution of the global mode from before the onset of vortex breakdown to a post breakdown state and found the wavemaker of the global mode to be located upstream of the recirculation bubble, from where initial $m = 1$ perturbations contaminated the entire flow field. Qadri et al. \cite{Qadri.2013} conducted a global stability analysis of a vortex breakdown bubble and identified the region of largest structural sensitivity in the vicinity of the upstream stagnation point. Tammisola \& Juniper \cite{Tammisola.2016} used DNS results of a swirl injector as the basis for their mean flow analysis. They show that the region of largest structural sensitivity and the wavemaker of the global helical mode is located upstream of the recirculation bubble. Consistently, a local analysis of the same injector yielded a wavemaker located at the upstream stagantion point \cite{Oberleithner.2015c}. Kuhn et al. \cite{Kuhn2016a} conducted lock-in experiments in a generic swirling jet by forcing the $m=1$ mode in the jet core. They found the lowest lock-in amplitudes right upstream the breakdown bubble indicating high structural sensitivity in this region.
\end{itemize} 
Considering these arguments, the subdominant saddle point \textbf{S2} is very likely to correspond to the limit-cylce oscillations we observe ion the experiments. It is worth mentioning that in recent investigations of coherent structures in a swirl combustor, at specific operating conditions, a second helical oscillatory mode with a much lower frequency  was observed \cite{Terhaar.2014b,Sieber2016b}. The frequencies of the first mode was in between $\mathrm{St}=0.5$ and $\mathrm{St}=0.6$ and for the second mode between $\mathrm{St}=0.1$ and $\mathrm{St}=0.2$, which is very similar to the frequencies of the dominant and subdominant saddle points. Although, the present data did not reveal any prominent second oscillatory mode, it is plausible that the saddle point found in the recirculation bubble is related to a mode that appears at slightly different operating condition. 
}
\FloatBarrier
\subsection{Stability Analysis of Model Profiles}
The purpose of the model study is twofold. First, we want to investigate, if the increased extent of the region of absolute instability of C1010 is particular to the one specific velocity and density field that we measured, or if this phenomenon persists in a general setup. Second, we want to investigate the influence of the relative position of the velocity and density gradient. This is motivated by the observations made in \figref{fig:VeloTMeas}, where the increase of the global mode amplitude was observed to be correlated to a collocation of the shear and temperature mixing layer. In addition, the importance of the relative position of the velocity and density gradient was discussed by Emerson et al. \cite{Emerson.2012} in the context of non-swirling bluff body stabilized combustion. These authors found that a collocation of the density and velocity gradient favors the transition to absolute instability. Before the results are presented, we discuss the modeling assumptions and the considered family of velocity and density profiles.\\
Numerous studies investigated the stability characteristics of jets, swirling jets and wakes \cite{Monkewitz.1988b,Gallaire.2003b,Jendoubi.1994,Michalke.1993,Michalke.1999}. It was shown that the stability of a swirling jet depends on many parameters, e.g. the Reynolds number, the shear layer thickness, the thickness of the temperature layer, the centerline velocity, the swirl intensity and the azimuthal wave number under consideration. To reduce the parameter space, the shear layer thicknesses of the axial and azimuthal velocity profile are fitted to velocity profiles that are obtained by averaging all PIV measurements indicated by the black circles in \figref{fig:KoKiEint} a). The model coefficients for the parameter study are extracted at $x/D = 0.5$. The details of the velocity model and a comparison of the velocity fit to measured data are available in \appref{sec:vModel}. The density profile was taken from the study of Yu \& Monkewitz \cite{Yu.1990}. To represent the measured profiles reasonably well, a shift of the radial coordinate was introduced (\appref{sec:vModel}). In addition, the azimuthal wave number was fixed to $m = 1$ and only the two Reynolds numbers of the experiment were considered. The centerline velocity $v_{\mathrm{cl}}$ and the swirl intensity parameter $q$ are the only free model parameters considered.  
%-----------------------------------------------------------------------------------------------------------------------------------------------------------------------------------------------
\subsubsection{The convective/absolute instability transition}
\begin{figure}
\centering
\includegraphics{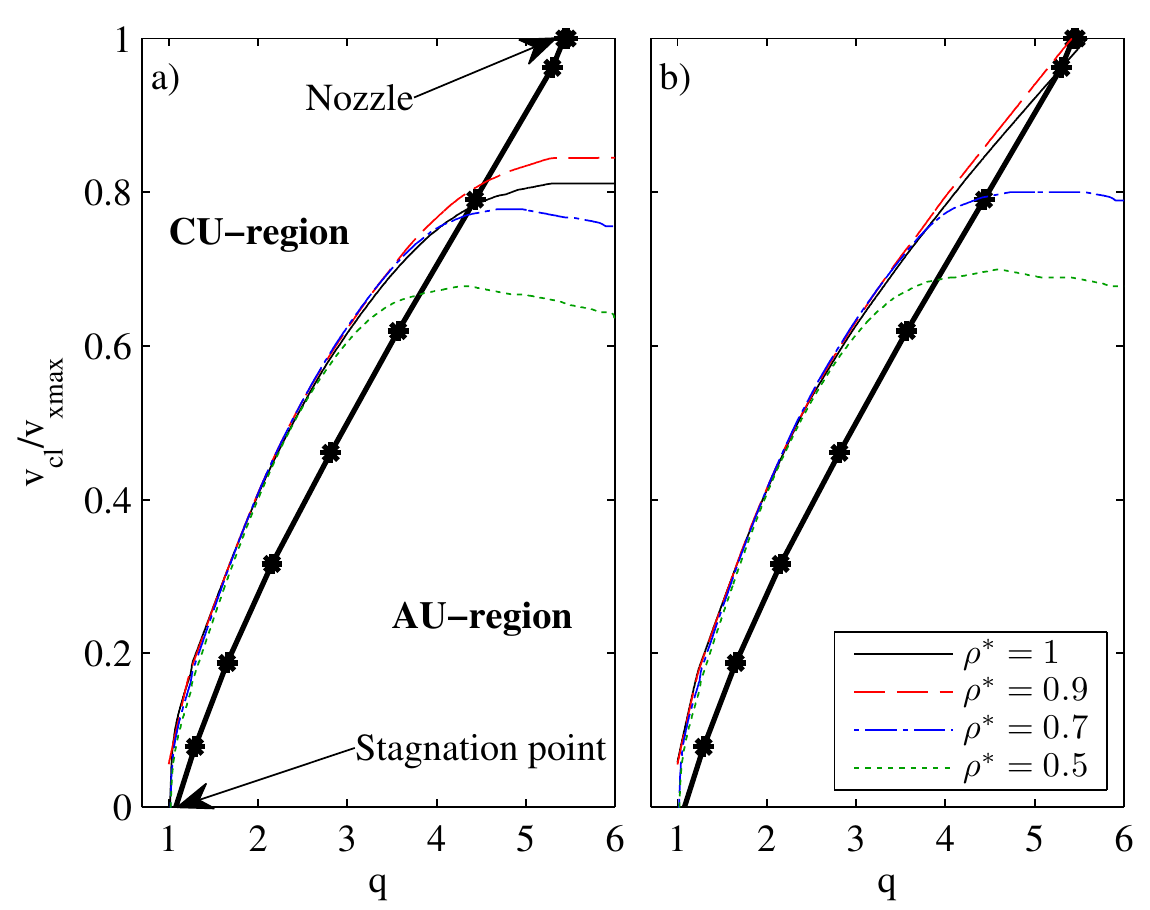}%
\caption{The boundary between convectively (CU) and absolutely (AU) unstable velocity profiles in the $q-v_{cl}$ parameter space at a Reynolds number of 4000 (a)) and 10000 (b)). The thick solid line with markers shows the $q-v_{cl}$ combinations of the measured velocity profiles when the $x$-axis is traversed from the nozzle to the stagnation point. The boundary from convective to absolute instability is indicated by the thin lines. The respective density ratios are indicated by the legend in b), which applies to a) and b). \label{fig:AU}}% }
\end{figure}

\Figref{fig:AU} a) and b) present the boundary between convective and absolute instability in the $v_{\mathrm{cl}}-q$ parameter space at a Reynolds number of 4000 and 10000. Each thin line corresponds to a different density ratio. The parameter combinations above each curve (towards larger values of $v_{\mathrm{cl}}$) yield convectively unstable velocity profiles. The thick line represents values of $q$ and $v_{\mathrm{cl}}$ derived from the calibration data set, showing one exemplary path through the parameter space, moving from the nozzle to the upstream stagnation point. The parameters in the top right corner of \figref{fig:AU} a) and b) correspond to velocity profiles in the vicinity of the nozzle, with larger values of the swirl and no velocity deficit on the jet center line. Values of $q$ and $v_{\mathrm{cl}}$ in the bottom left corner of the figure correspond to velocity profiles in the vicinity of upstream stagnation point. The swirl has decayed and the wake depth has increased.    \\
A comparison of \figref{fig:AU} a) and b) shows that the influence of viscosity is not decisive for the transition from convective to absolute instability. For each density ratio, the larger Reynolds number is linked to a larger parameter space that yields absolutely unstable velocity profiles. Qualitatively, the transition from convective to absolute instability is not influenced by the Reynolds number. The difference in Reynolds number between C10YY and C25YY therefore provides no explanation for the different behavior of the global mode that we observed so far.\\ 
Considering the boundary between convective and absolute instability for the different density ratios, we recover the influence of swirl on the transition from convective to absolute instability at $\rho^{*} = 1$. The larger the value of the swirl parameter, the smaller the velocity deficit necessary to obtain absolute instability \cite{Gallaire.2003b}. This observation persists at all density ratios considered.
In the parameter range of $q$ from 1 to 2.5 and $v_{\mathrm{cl}}$ from 0 to 0.5, the transition from convective to absolute instability is not influenced by heating (\figref{fig:AU} a) and b)). The shear instability induced by the deep wake is dominating over the impact of density stratification in this regime. This observation is consistent with \figref{fig:Absamp10} and \figref{fig:Absamp25}, where the heating had little impact on the absolute growth rate for $x/D > 0.6$. A strong change in the position of the convective/absolute instability boundary is evident in \figref{fig:AU} a) and b) for the remaining parameter space and particularly for parameter combinations that put the resulting velocity profile in the vicinity of the nozzle. Mild heating increases the absolutely unstable parameter space in this region: Compared to isothermal conditions a smaller velocity deficit is sufficient to trigger absolute instability, when mild heating is applied. This is the case for C1010, where the transition point from convective to absolute transition is at the most upstream location of all configurations of C10YY.
%-----------------------------------------------------------------------------------------------------------------------------------------------------------------------------------------------
\subsubsection{The impact of density collocation}  
\begin{figure}
\centering
\includegraphics{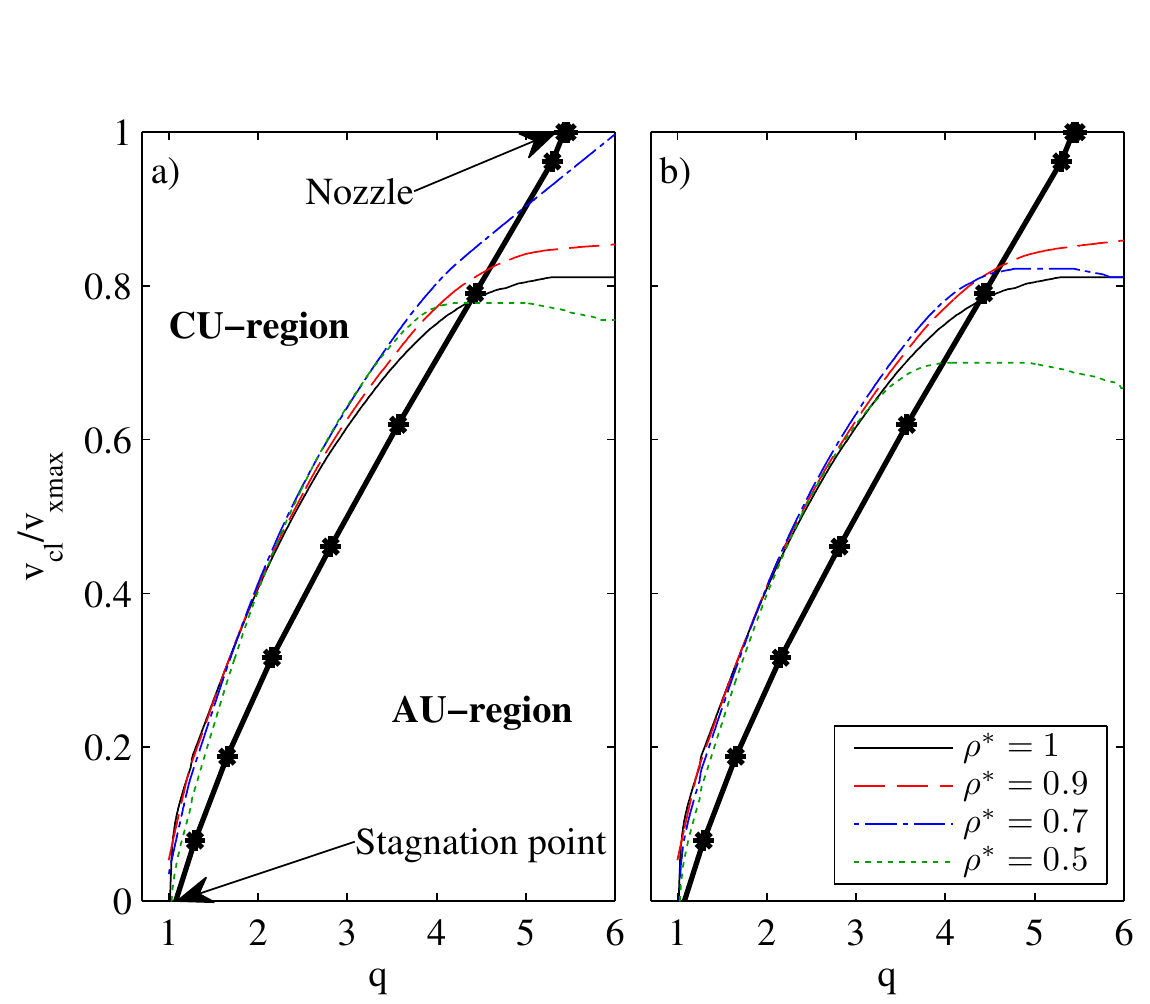}%
\caption{The boundary between convective (CU) and absolute (AU) instability for a relative shift of the density profile of 0.1D (a)) and 0.2D (b)). Other than that, the caption of \figref{fig:AU} applies.\label{fig:AUtlS}}%
\end{figure}

Recalling that we observed different relative positions of the inner shear layer and the temperature mixing layer in \figref{fig:VeloTMeas}, we now consider how a shift of the density profile relative to the velocity profile influences the transition to absolute instability. Relative to the baseline shift, the density model (\appref{sec:vModel}), \figref{fig:AU} a) is recomputed for an outward radial shift by 0.1D and 0.2D. Note that a shift of 0.1D is present in the experimental data (\figref{fig:VeloTMeas}). From \figref{fig:AUtlS} a) and b) it can be inferred that a shift of the density profile by a small amount significantly influences the transition between absolute and convective instability. A strong increase in the absolutely unstable parameter space is produced by a density profile shift of 0.1D (\figref{fig:AUtlS} a)). Absolute instability is promoted by density ratios as small as 0.7 compared to isothermal conditions. Shifting the density profile further outward by 0.2D (\figref{fig:AUtlS} b)) further promotes the onset of absolute instability by density ratios down to 0.7. The overall extent of absolutely unstable parameter combinations is reduced compared to the situation in \figref{fig:AUtlS} a).\\
How does this relate to experimental observations? \Figref{fig:VeloTMeas} outlines that the relative position of the inner shear layer and the temperature mixing layer varies along the axial direction and by as much as 0.1D between different configurations. The model study confirms that a shift of 0.1D drastically influences the onset of absolute instability and that the onset of absolute instability may be triggered early on by heating the flow, before a strong density gradient strongly reduces the absolutely unstable parameter space.      
\textcolor{black}{We note that these observations are in line with the studies of Lingens et al. \cite{Lingens.1996} and Nichols et al. \cite{Nichols.2007}. Lingens et al. investigated the onset of absolute instability in a round jet diffusion flame. Their experiment consisted of a fuel jet and a coanullar oxidizer stream that exited into the ambient. The flow field in the experiment of Lingens et al. is similar to the flow field in the present investigation in that it also features two axial shear layers. Additionally, the density gradient is aligned with the shear layer between the oxidizer and the fuel stream (the inner shear layer in this investigation). Lingens et al. report that the modification of the absolute instability is linked to the shear layer between fuel and oxidizer, which reamphasizes the importance of the collocation of the inner shear layer and the mixing layer in this study.\\
Nichols et al. \cite{Nichols.2007} consider self-sustained oscillations in variable density round jets without swirl. They find that the growth rate of the absolute instability is strongly related to the collocation of the velocity and the density profile. These authors find that the growth rate of the absolute instability increases with a stronger collocation of the shear layer and the mixing layer. This is consistent with our observations that the global mode initially increases in amplitude.}    
%-------------------------------------------------------------------------------------------------------------------------------------------------------------------------------------------------------------------
%-------------------------------------------------------------------------------------------------------------------------------------------------------------------------------------------------------------------
\FloatBarrier
\section{Conclusion}
In this study we investigated the evolution of the global mode in unconfined heated swirling jets undergoing vortex breakdown. In the experimental setup, heating was applied exclusively to the breakdown bubble, which is similar to certain configurations found in swirl stabilized combustion. The aim of this study was to investigate the density-driven suppression of the global mode in a well-controlled environment. The measured flow velocity and temperature fields reveal that the amplitude reduction of the global mode is not an "on-off'' transition, but many intermediate states exist. To our surprise, we found that mild heating of the breakdown bubble leads to an increase in the amplitude of the global mode, which has not been observed so far. Strong heating, however, leads to the suppression of the global mode, which confirms the previous observations in reacting swirling jets. We applied local linear stability analysis to the measured mean flow and density field and determined the wavemaker and selected global frequency. The agreement with experiment is excellent for all considered configurations, which justifies the analytic approach, \textit{a posteriori}. The vanishing growth rate of the global mode further indicates that the nonlinear saturation of the global mode at its limit cycle is well represented by the mean flow. The concept of the global mode wavemaker is particularly useful, as it relates the suppression/excitation of the global mode to a single point. We showed that the density ratio at the wavemaker location is indeed decisive for the suppression/excitation of the global mode, rather than the overall density ratio.\\
To generalize these results, local linear stability analysis was applied to model velocity and density profiles. The model study successfully captured the qualitative change of absolute instability with mild and strong heating. This effect was demonstrated to be independent of the Reynolds number, but strongly dependent on the collocation of the density profile and the inner shear layer. The increase or decline of the global mode is clearly related to two factors: The density ratio at the wavemaker of the flow and the collocation of the density and velocity profile in the inner shear layer. Depending on the collocation and the density ratio, the global mode is suppressed or excited. \\
The analytic framework presented in this study is well suited for more complex flow configurations, as encountered in turbulent swirl-stabilized combustion. The results from this fundamental study are particularly relevant for future flow control applications in this area. The knowledge of the wavemaker position and of the importance of the velocity and density collocation lay out the path for control strategies that directly act on the source of the instability.
%-----------------------------------------------------------------------------------------------------------------------------------------------------------------------------------------------
% If you have acknowledgments, this puts in the proper section head.
\begin{acknowledgments}
The financial support from the German Science Foundation (DFG) under project grants PA 920/29-1 and PA 920/30-1 is gratefully acknowledged.
\end{acknowledgments}
%-----------------------------------------------------------------------------------------------------------------------------------------------------------------------------------------------
\begin{appendix}

\FloatBarrier
\section{The influence of the heating element on the flow field}
\label{sec:HE}
\begin{figure}
\includegraphics{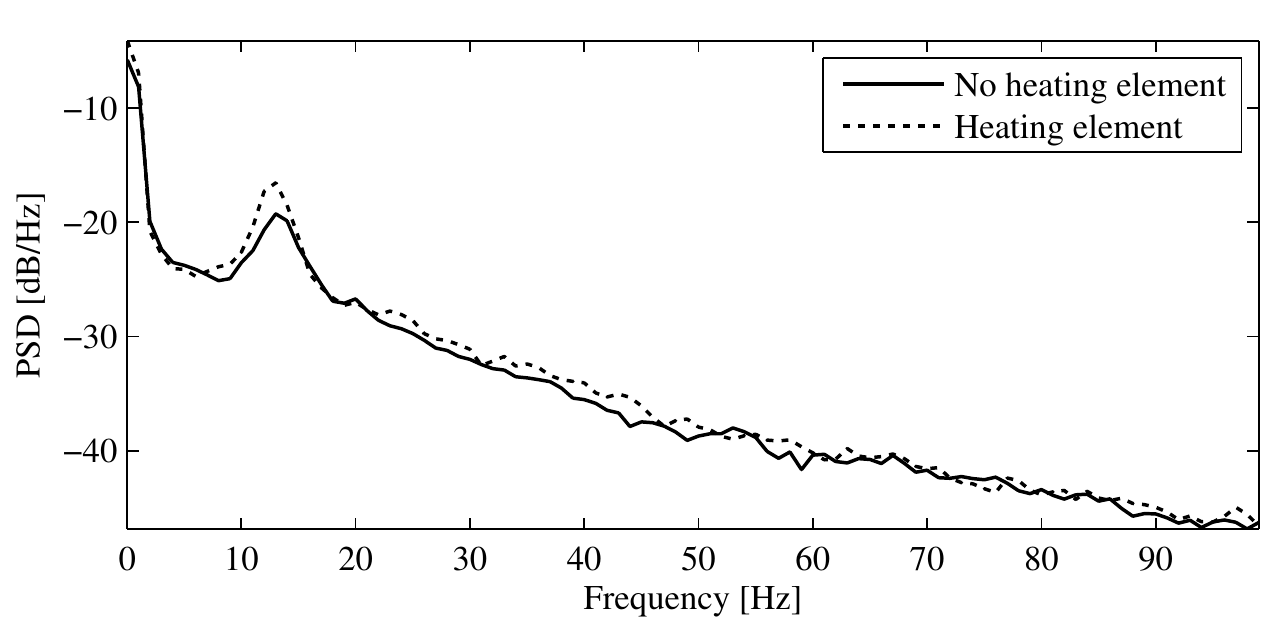}%
\caption{Frequency spectra obtained with a hotwire for the baseline case and with the heating element present.\label{fig:fHE}}%
\end{figure}

\begin{figure}
\includegraphics{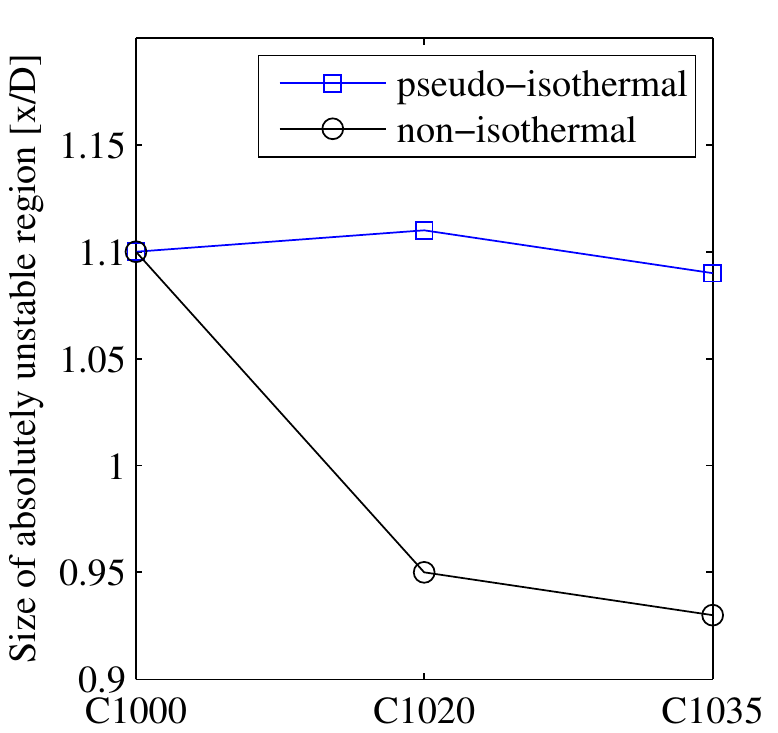}%
\caption{\textcolor{black}{The size of the region of absolute instability for selected flow configurations.}\label{fig:auSize}}%
\end{figure}

It is well known that the placement of a control cylinder may have a profound impact on the formation of the global mode in cylinder wakes \cite{Strykowski.1990,GIANNETTI.2007}. Therefore, it has to be verified, whether the heating element alone has any impact on the structure or energy of the global mode.
\begin{figure}
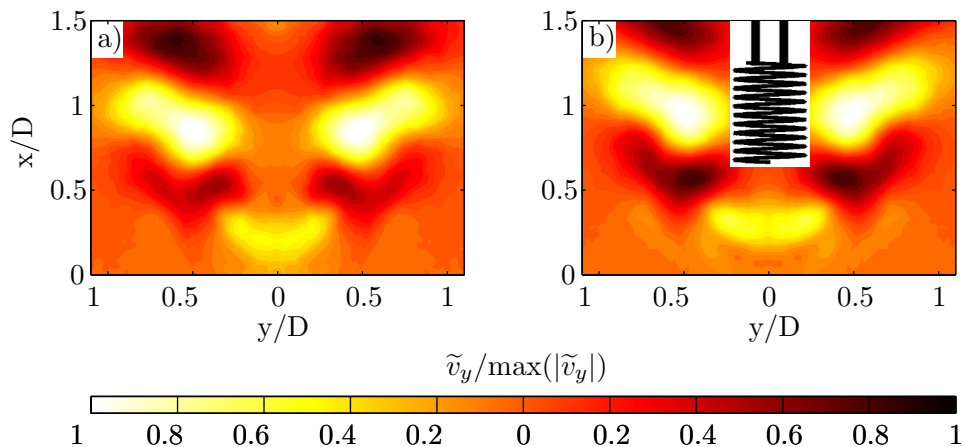
\caption{Spatial structure of the global mode. a) Without the presence of a heating element. b) With the presence of the heating element and $\rho^{*} = 1$.\label{fig:HE}}%
\end{figure}
The influence of the heating element is studied with the help of \autoref{fig:HE}. Different spacings between the upstream end of the heating element and the nozzle were investigated. The results shown here are for the smallest distance between the nozzle and the heating element, with the upstream end of the heating element positioned at $x/\mathrm{D} = 0.6$. The influence of the heating element decreases, when the heating element is positioned further downstream. \Figref{fig:HE} a) and b) show the spatial structure of the global mode derived from a POD analysis of the PIV measurements. The baseline configuration is presented in \figref{fig:HE} a). In this case the heating element is not present in the flow. \Figref{fig:HE} b)  shows the spatial structure of the global mode in the presence of the heating element. Note that the heating element is positioned in the flow, but no heating current was supplied. As it is evident from these two plots, the spatial structure and axial wavelength remain virtually unchanged. \\
%From the time-averaged axial velocity it becomes evident that the presence of the heating element increases the backflow velocity in the recirculation bubble. We note further that the POD indicates a 3\% increase in contribution to the total kinetic energy budget, if the heating element is present.\\
The temporal dynamics of the global mode were assessed with a hotwire positioned in the outer shear layer at $y/\mathrm{D} = 0.55$ and $x/\mathrm{D} = 0.3$. At this location, a good signal to noise ratio between the frequency of the global mode and turbulent fluctuations was obtained. Comparing the two curves shown \figref{fig:fHE}, it can be asserted that the presence of the heating element only slightly shifts the frequency of the global mode to higher values and  that the energy contained in the flow at the global mode frequency also only slightly increases. We conclude that the presence of the heating element alone has no significant effect on the global mode. Specifically, the heating element does not suppress the global mode. \\
\textcolor{black}{Besides the passive impact of the heating element, it is also important  to study the influence of the heat-addition on the mean flow and its stability. 
For that reason, stability analysis is conducted for a pseudo isothermal flow, by taking the mean flow of the heated jet and setting the density constant. Results are compared to the analysis based on the same mean flow and the actual density field. \\
We take the streamwise extent of the region of absolute instability as a good representative of the overall stability characteristics and compare its evolution for different heating powers for configuration C1000, C1020 and C1035 in \figref{fig:auSize}. Evidently, the size of the region of absolute instability remains practically constant for the pseudo-isothermal flow fields. This indicates that the heated flow is not additionally influenced by the presence of the heating element. The black line shows the evolution of the size of the region of absolute instability for the analysis based on the nonuniform flow and density field. In contrast to the pseudo-isothermal flow the region of absolute instability is substantially altered in its size. This indicates that the change in flow stability is indeed related to the change of the density field and not to a potential change of the mean flow filed due to the heating.}
%----------------------------------------------------------------------------------------------------------------
\FloatBarrier
\section{The velocity model}
\label{sec:vModel}
\begin{figure}
\includegraphics{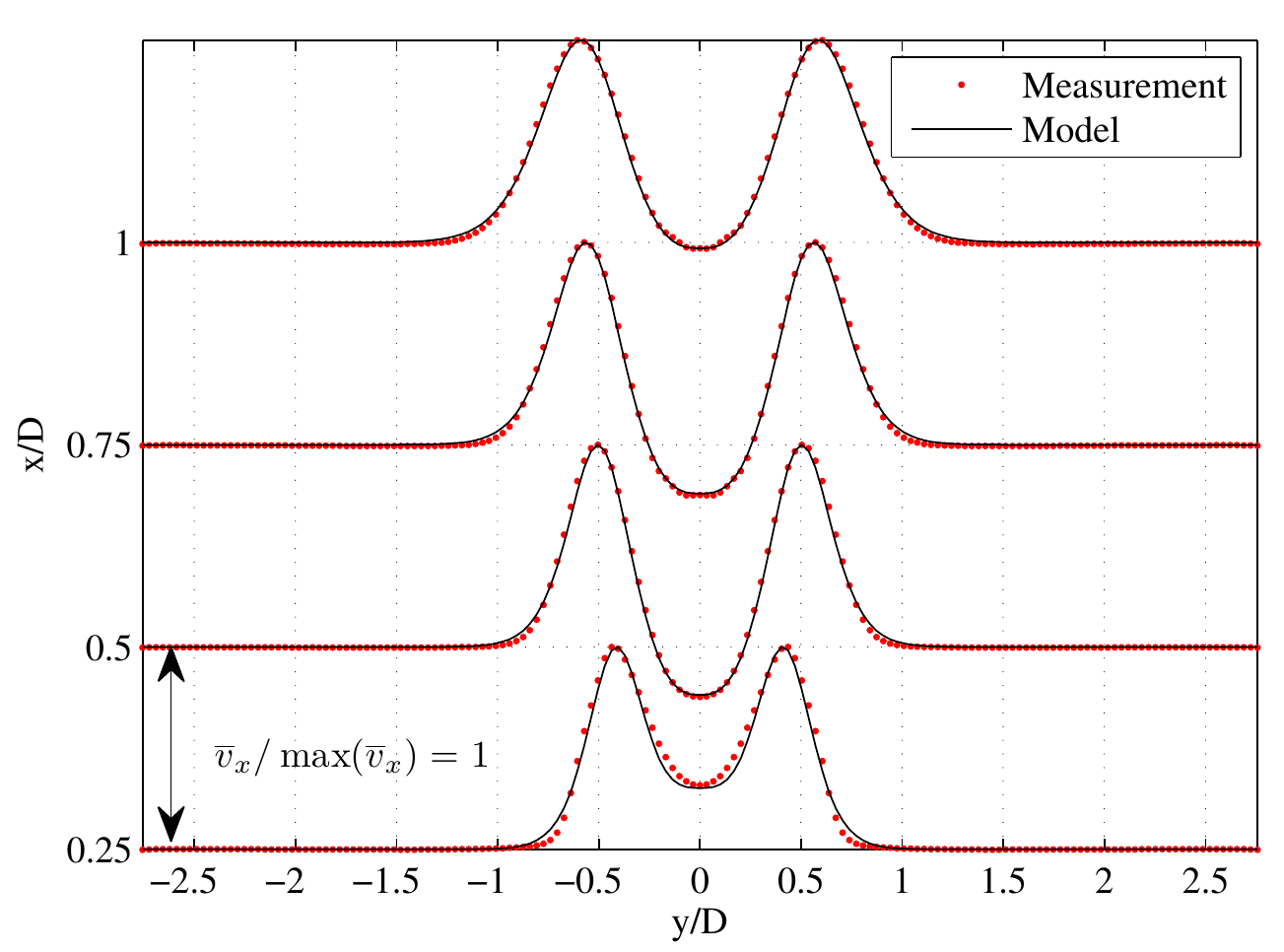}%
\caption{Measured and modeled axial velocity profiles.\label{fig:axialVeloModel}}%
\end{figure}

\begin{figure}
\includegraphics{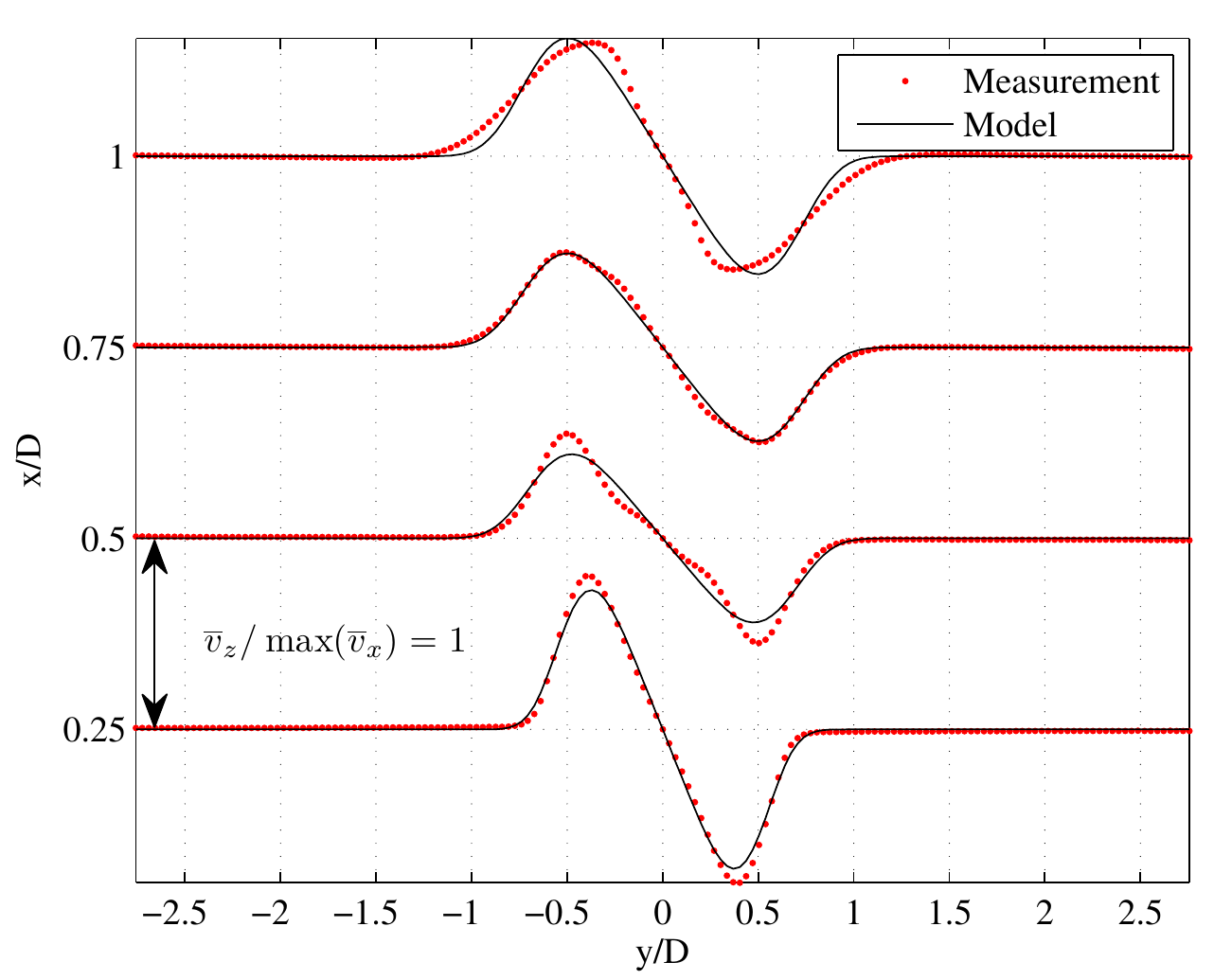}%
\caption{Measured and modeled out-of-plane velocity profiles.\label{fig:azimVeloModel}}%
\end{figure}

A fictitious velocity field is obtained by averaging all velocity fields indicated by circles in \figref{fig:KoKiEint} a), in order to avoid any particularities of Q1010. 
The axial and azimuthal velocity are modeled with the functions
\begin{align}
\label{eq:axmod}
v_{x}(r) &= 4\cdot B \cdot F_{x1}\left(1 - B\cdot F_{x2}\right)\\
B &= 0.5\cdot \sqrt{1 - v_{\mathrm{cl}}}\\
\intertext{and}
\label{eq:azmod}
v_{\theta}(r) &= q \cdot r \cdot \left(F_{\theta1} + F_{\theta2}\right),\\
\intertext{with}
F_{x1} &= \frac{1}{1+\left(\exp\left(\ts{c}{v_{xmax}}\cdot r^{2}\right)-1\right)^{\ts{c}{\delta1}}}\\
F_{x2} &= \frac{1}{1+\left(\exp\left(\ts{c}{v_{xmax}}\cdot r^{2}\right)-1\right)^{\ts{c}{\delta2}}}\\
F_{\theta1} &= \ts{c}{\theta1} \cdot \exp\left(-\left(r/\ts{c}{v_{\theta max}}\right)^{4}\right)\\
F_{\theta2} &= \ts{c}{\theta2} \cdot \exp\left(-\left(r/\ts{c}{v_{\theta max}}\right)^{2}\right).
\end{align}
The velocity model for the axial component (\autoref{eq:axmod}) was originally introduced by Michalke \cite{Michalke.1999}. The coefficients $\ts{c}{v_{xmax}}$ determine the radial position of the maximum of the axial velocity profile. The thickness of the inner and outer shear layer is controlled by the shape parameters $\ts{c}{\delta1}$ and $\ts{c}{\delta2}$. The model for the azimuthal velocity component was inspired by the work of Gallaire \& Chomaz \cite{Gallaire.2003b}. $\ts{c}{v_{\theta max}}$ controls the radial position of the maximum of the azimuthal velocity component and $\ts{c}{\theta1}$ and $\ts{c}{\theta2}$ scale the magnitude of $F_{\theta1}$ and $F_{\theta2}$. This allows for continuously changing the slope of the azimuthal velocity profile across the jet centerline.
The value of the coefficients derived from a model fit to the fictitious data are listed in \autoref{tab:Coef}. The two remaining parameters in the velocity profiles $v_{\mathrm{cl}}$ and $q$ quantify the velocity on the centerline and the swirl intensity. The swirl parameter $q$ is defined as
\begin{equation}
q = \frac{\Omega_{c}\ts{r}{sl}}{\mathrm{max}(v_{x})},
\end{equation}
where $\Omega_{c}$ is the rotation rate of the vortex core on the jet axis, $\ts{r}{sl}$ the position of the outer shear layer and $\mathrm{max}(v_{x})$ the maximum value of the axial velocity at each flow slice. \Figref{fig:axialVeloModel} and \figref{fig:azimVeloModel} compare the model fits to the calibration data at various axial positions.\\
\begin{table}
\caption{Velocity and density model parameters.\label{tab:Coef}}     
\begin{tabular}{c c c c c c c}
\toprule
 $\ts{c}{v_{xmax}}$ & $\ts{c}{\delta1}$ & $\ts{c}{\delta2}$ & $\ts{c}{v_{\theta max}}$ & $\ts{c}{\theta1}$ & $\ts{c}{\theta2}$ & \ts{c}{\rho} \\
\hline
2.36 & 1.36 & 2.34 & 0.67 & 1.18 & 0 & 1.4\\
\botrule
\end{tabular}
\end{table}

\begin{figure}
\includegraphics{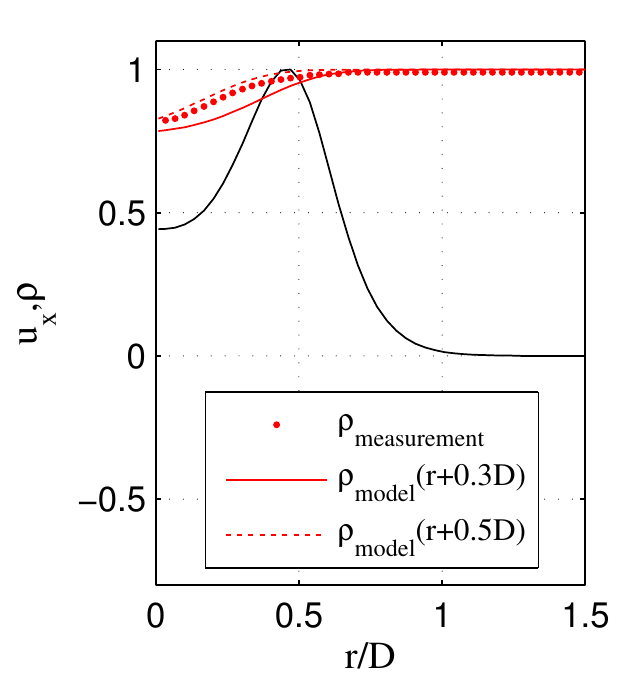}%
\caption{The dotted red line indicates one density profile obtained from measurement. The dashed and solid red line indicate the model density profiles with $r_{\rho} = 0.5D$ and $r_{\rho} = 0.3D$, respectively. An exemplary axial velocity profile is given for reference.\label{fig:rhoModel}}%
\end{figure}

%---------------------------------------------------------------------------------------------------------------------------------------------------------------------------------------------
Similar to Yu \& Monkewitz \cite{Yu.1990}, the density profile is assumed to be of the form
\begin{align}
\rho(r) &= 1- \left(\left(1 - \rho^{*} \right) \cdot F_{\rho}\right),\label{eqn:dModel}\\
\intertext{with, in this study,} 
F_{\rho} &= \exp\left(-\left(c_{\rho} \cdot (r+r_{\rho}) \right)^{4}\right).
\end{align}
$\ts{c}{\rho}$ determines the radial position, at which the density is equal to ambient conditions. The density profile given by \eqnref{eqn:dModel} models the data well with a shift of $r_{\rho} = 0.5D$. We consider this as our baseline configuration. \Figref{fig:rhoModel} illustrates different density profiles, where the dotted red line represents a density profile obtained from experiment. The dashed red line is the model profile with $r_{\rho} = 0.5$D (the baseline configuration). The solid red line shows the model density profile with $r_{\rho} = 0.3$D, which is the largest shift in the radial outward direction we consider here. Note that in the discussion of the model study results, the configurations are always given in relation to the baseline case, so that the configuration with an absolute shift of $r_{\rho} = 0.3$D has a relative shift of 0.2D to the baseline case with $r_{\rho} = 0.5$D.\\
In studies on hot jets, wakes and mixing layers \cite{Monkewitz.1988,Yu.1990,Khorrami.1995,Lu.1999,Lesshafft.2007,Srinivasan.2010}, the density profile was computed directly from the axial velocity profile under the Crocco-Busemann relation, which is exactly valid for a Prandtl number of one. In the present study, the axial velocity profile would be given by $F_{\rho}$, when $r_{\rho} = 0$. Such a model inadequately describes the measurements in this study. Lesshafft \& Marquet \cite{Lesshafft.2010} investigated hot jets and pointed out that the direct computation of the density profile from the velocity profile via the Crocco-Busemann relation is inadequate for determining a general upper bound for the critical density ratio of absolutely unstable hot jets. They found a value of 0.9 for the transition to absolute instability, in contrast to the value of 0.72 found by Monkewitz \& Sohn \cite{Monkewitz.1988}. Considering this result and the inability of the Crocco-Busemann relation to capture the measured velocity and density profiles in this study, together with the profound importance of the collocation of the inner shear layer and the temperature mixing layer, we suggest that the velocity and density profiles in swirling jets need to be modeled separately. Otherwise, the stability properties of heated swirling jets may not be captured correctly.
%%----------------------------------------------------------------------------------------------------------------
%\FloatBarrier
%\textcolor{red}{\section{The quality of the Pad\'{e} polynomial fit}
%\label{sec:fitQuali}
%The quality of the fit to the absolute growth rate curve is critical for the accuracy of the wavemaker determination. The accuracy of the fit is shown for C2560 in \figref{fig:fitQuali}, a similar fit quality is achieved for all other cases as well. Evidently, the quality of the fit is very good.
%\begin{figure}
%\includegraphics{../Bilder/fitQual.pdf}%
%\caption{The curves of $\Im{\omega_{0}}$ (left) and $\Im{\omega_{0}}$ (right), together with the polynomial fit to each curve for C2560. The solid line indicates $\omega_{0}$ and the circles mark the fit.\label{fig:fitQuali}}%
%\end{figure}}
%\FloatBarrier
\end{appendix}
%-----------------------------------------------------------------------------------------------------------------------------------------------------------------------------------------------

\bibliography{pof}
% Create the reference section using BibTeX:

\end{document}